\documentclass[aps,preprint]{revtex4}

\usepackage[dvips]{graphicx}

\newcommand{\ch}{{\cal H}}
\newcommand{\ce}{{\cal E}}

\begin{document}
\date{\today}

\title{Nearby states in non-Hermitian quantum systems}

\author{Hichem Eleuch$^{1}$\footnote{email: 
heleuch@fulbrightmail.org} and  
Ingrid Rotter$^{2}$\footnote{email: rotter@pks.mpg.de, author to whom
  correspondence should be addressed}}

\address{
$^1$
Department of Physics, McGill University, Montreal, Canada H3A 2T8}
\address{
$^2$Max Planck Institute for the Physics of Complex Systems,
D-01187 Dresden, Germany }

\begin{abstract}

In part I, the formalism for the description of open quantum systems
(that are embedded  into a common well-defined environment) by means of 
a non-Hermitian Hamilton operator $\ch$ is sketched. Eigenvalues and 
eigenfunctions are parametrically controlled. Using a 2$\times$2 model, 
we study the eigenfunctions of $\ch$ at and near to the singular 
exceptional points (EPs) at which two eigenvalues coalesce and the 
corresponding eigenfunctions differ 
from one another by only a phase. In part II,
we provide the results of an analytical study for the 
eigenvalues of three crossing states. These crossing points are of
measure zero. Then we show numerical results 
for the influence of a nearby ("third") state onto an EP. 
Since the wavefunctions of the two crossing states are mixed in a finite 
parameter range around an EP, three states of a physical system
will never cross in one point.
Instead, the wavefunctions of all three states are mixed in a finite 
parameter range in which the ranges of the influence of different EPs
overlap. We may relate these results  to  dynamical phase 
transitions observed recently in different experimental studies. 
The states on both sides 
of the phase transition are non-analytically connected.

\end{abstract}

\maketitle

\section*{\large \bf Part I: Two states}

\section*{Abstract}

The formalism for the description of open quantum systems (that are 
embedded  into a common well-defined environment) by means of 
a non-Hermitian Hamilton operator $\ch$ is sketched. Eigenvalues and 
eigenfunctions are parametrically controlled. Using a 2$\times$2 model, 
we study the eigenfunctions of $\ch$ at and near to the singular 
exceptional points (EPs)
at which two eigenvalues coalesce and the corresponding eigenfunctions differ 
from one another by only a phase. Nonlinear terms in the Schr\"odinger 
equation appear nearby EPs which cause a mixing of the wavefunctions in a 
certain finite parameter range around the EP. The phases of the eigenfunctions
jump by $\pi$ at an EP. These results hold true for systems that 
can emit ("loss") particles into the environment of scattering
wavefunctions as well as for systems which can moreover 
absorb ("gain") particles from the environment. In a parameter range far 
from an EP, open quantum systems are described well by a
Hermitian Hamilton operator. The transition from this parameter range to
that near to an EP occurs smoothly. 

%\end{abstract}

%\pacs{\bf 03.65.Ta;   03.65.Ca;   %03.67.Bg}
\maketitle

\section{Introduction}
\label{intr}

The basic features of quantum mechanics are worked out  about
90 years ago: the Schr\"odinger equation  is linear
and allows superpositions of quantum states to be solutions 
of the Schr\"odinger equation;
the Hamiltonian $H^B$ describing the system is Hermitian, 
its eigenvalues $E_i^B$ are real and its
eigenfunctions $\Phi_i^B$ are normalized according to 
$\langle\Phi_i^B|\Phi_j^B\rangle = \delta_{i,j}$. 
The system described in this manner  is  {\it closed} since its
coupling to an environment is not  involved in the theory. 
The finite lifetime of most states of a (small) system 
is calculated by means of tunneling, without taking into account 
any feedback from the environment
onto the system. This theory is proven experimentally 
during multi-year  studies performed on different systems at low level
density. 

For the last years, not only the resolution of most
experimental devices has increased considerably 
but also calculations with higher accuracy have become possible.  
As a result, the standard quantum theory has shown its limit to
describe successfully  experimental results. Counterintuitive results
are obtained in different experiments. An example is the observation
of an unexpected regularity of the measured transmission phases
(so-called phase lapses) in mesoscopic systems \cite{heibl}
which could not explained in the framework of Hermitian quantum
physics in spite of much effort \cite{focus1,focus2}. They are explainable 
however by considering the feedback from the environment
onto the system \cite{muro}. Another example is the experimental
observation and theoretical description
of a {\it dynamical phase transition (DPT)} in the spin swapping 
operation \cite{past1,past2}. While Fermi's golden rule holds below the DPT, 
it is violated above it. In a new experimental
paper \cite{bird14a}, the formation
of a protected sub-band for conduction in quantum point contacts under
extreme biasing is found, see also \cite{bird14b}.  
This sub-band is a collective robust mode of 
non-equilibrium transport that is immune to local heating. It has
potential practical implications for nanoscale devices made of quantum
point contacts and quantum dots. 

In order to improve the theoretical description, in some papers
the coupling of the system to an environment is taken into account
explicitly. Mostly, this is done by replacing the Hermitian
Hamilton operator, or part of it, by a non-Hermitian one, see
e.g. the reviews \cite{ro91,top} and the book \cite{mois}. 
In other papers, nonlinearities are added to the Schr\"odinger
equation. An example is the review   \cite{flach} where the role of 
nonlinear Fano resonances in theoretical and
experimental studies of light propagation in photonic devices and charge
transport through quantum dots (nanostructures) is reviewed. 
By this means, the description of experimental results could be improved
considerably in all cases.

A non-Hermitian Hamiltonian in the Schr\"odinger equation appears when 
the system is considered to be {\it open}, i.e. to be
embedded into an environment, and the coupling between the
system and its environment is taken into account from the very beginning. 
A natural environment is the continuum
of scattering wavefunctions to which the states of the system 
are coupled and into which they decay.  
It can be changed by external fields, however never be deleted. 
The finite lifetime of the states of the system
is calculated directly from the non-Hermitian part of the
Hamiltonian \cite{top,ro91}. The feedback from the 
environment onto the system is involved in the non-Hermitian
Hamiltonian $\ch$ and therefore also in its eigenvalues $\ce_i$ and
eigenfunctions $\Phi_i$. The basic assumption of this description 
is supported experimentally by the recent  observation that
remote states are coupled through the continuum \cite{birdprx}.

Meanwhile there are many calculations performed with a
  non-Hermitian Hamiltonian. Usually, the behavior of the system is
  controlled by means of varying a certain parameter.
The restriction of the parameter dependence  of the
Hamiltonian $\ch $ to its explicitly non-Hermitian part (by neglecting
the parameter dependence of its real Hermitian part) allows us to receive a
quick overview on the spectroscopic redistribution processes occurring in the
system under the influence of the coupling to the environment, 
see e.g. \cite{jumuro,muro,celardo2}. Most interesting is the
appearance of unexpected collective coherent phenomena  in different 
systems. They are similar to the phenomenon of Dicke superradiance \cite{dicke} 
which is known in optics for many years.  It has been shown, moreover, 
that the reorganization of the spectrum of the system under the
influence of the coupling to the environment at a
critical value of the control parameter,  occurs globally over the 
whole energy range of the spectrum \cite{jumuro}. It takes place by a 
{\it cooperative action of all states}, and  the length scale
diverges as well as the degree of non-Hermiticity of the Hamiltonian.
It has been shown further that the reordering of the spectrum 
corresponds, indeed,  to a second-order
phase transition \cite{jumuro}, justifying the  notation {\it dynamical
  phase transition}. The states below and beyond the DPT are
non-analytically connected. This method is shown to describe 
also phase transitions in, e.g., biological systems \cite{celardo1}.

The calculation of the eigenvalues $\ce_i$ and
eigenfunctions $\Phi_i$ of the non-Hermitian Hamiltonian $\ch$
hits upon some mathematically non-trivial problems due to the
existence of  singular points in the continuum. At these points,
two eigenvalues coalesce and the two corresponding eigenfunctions differ
from one another only by a phase \cite{top,comment1}. 
The geometric phase of these points
differs from the Berry phase of a diabolic point
by  a factor 2. These singular points, called usually {\it
  exceptional points (EPs)}, are well-known in mathematics
\cite{kato}. Their meaning for the dynamics of open quantum systems 
and the behavior of the two eigenfunctions  at an EP is
however studied only recently.
Numerical results for the eigenvalues and eigenfunctions of $\ch$ 
under the influence of an EP in a
concrete system  are obtained, e.g., for atoms \cite{marost1,marost2}, 
for the transmission through quantum dots \cite{burosa1,burosa2,rosabic,rosa} and
for charge transport in molecular networks \cite{peskin}. In the early
papers, the EPs are called mostly {\it branch points in the complex plane}
or {\it double poles of the S matrix}. 
Phase transitions in open quantum systems which are
associated with the formation of long-lived and short-lived states
according to  \cite{jumuro}, are related to EPs first in \cite{hemuro}. 
More recent results can be found in the review \cite{top}.
The drawback of all these studies is the unsolved question 
how different EPs influence one another and how they are related to a DPT.

The eigenfunctions of a symmetric non-Hermitian operator ${\cal H}$ are biorthogonal
according to $\langle \Phi_i^*| {\cal H }= \ce_i \langle \Phi_i^*|$ and 
${\cal H } |\Phi_i\rangle = \ce_i | \Phi_i\rangle$ (where $\ce_i$ is
a complex eigenvalue of  ${\cal H }$). They have to be normalized
therefore by means of $\langle \Phi_i^* | \Phi_j \rangle$ which is a
complex number (in difference to the norm  $\langle \Phi_i | \Phi_j
\rangle$ which is a real number). In order to guarantee a smooth
transition from the description of an open quantum system to an 
almost (and eventually really) closed one, the eigenfunctions of $\ch$ should be
normalized according to   $\langle \Phi_i^* | \Phi_j \rangle =\delta_{ij}$. This
is possible only by the additional requirement
Im$\langle \Phi_i^* | \Phi_j \rangle =0$. This condition implies
that the relation between the phases of the two states $i$ and $j$ is,
generally, not rigid:  far from an EP, the two wavefunctions are (almost)
orthogonal to one another in (nearly) the same manner as the eigenfunctions of
a Hermitian operator while they become  linearly dependent in 
approaching an EP \cite{top} such that the biorthogonality of them
cannot be neglected. This is quantitatively expressed by 
the {\it phase rigidity}  $r_i \equiv     
{\langle \Phi_i^* | \Phi_j\rangle}/{\langle \Phi_i |
  \Phi_j\rangle}$ which is reduced in approaching an EP, $r_i \to 0$. 
Here, the environment can put its information into the system 
by aligning  states of the system with  states of the environment,
i.e. by enhancing their decay width. 

The phase rigidity of the eigenfunctions and its reduction near to the
singular EP is the most interesting value when a realistic quantum system is 
described by a Schr\"odinger equation with non-Hermitian Hamiltonian. 
Since the environment is  able  to  change the spectroscopic
properties of the system only if  $r_i <1$,  an EP may 
influence strongly the dynamics of an open quantum system.  
This is in contrast to a closed system described 
by a Hermitian operator and rigid phases ($r_i =1$) of its eigenfunctions.
In \cite{burosa1,burosa2}, the correlation between 
non-rigid phases of the eigenfunctions  $\Phi_i$ of the non-Hermitian 
Hamiltonian  in the neighborhood of an EP and the
transmission through a quantum dot is demonstrated in 
calculations for a special quantum dot. The enhancement is a
collective effect caused by  $r_i <1$ for many levels $i$ in a certain
finite parameter range. It has been shown further \cite{top} that the Schr\"odinger
equation of the system contains nonlinear terms when  $r_i <1$, i.e. in the
  neighborhood of EPs. In contrast to the usual calculations, it is
 therefore not necessary to introduce nonlinear terms into the
 Schr\"odinger equation by hand. They are part and parcel of the
 non-Hermitian quantum physics, and appear {\it only} in the vicinity 
of EPs (where $r_i <1$).

Recently,  non-Hermitian Hamiltonians are studied  the
eigenvalues of which are real in a broad parameter range \cite{bender}.
Under certain conditions, the eigenvalues of the Hamiltonian  become
complex as shown theoretically \cite{bender} as well as experimentally 
\cite{mumbai1,mumbai2,mumbai3,schindler}. The meaning of EPs for these 
processes is studied in different papers, e.g. 
\cite{opt,elro4a,berggren,joglekar,peng1,peng2,peng3}. 
Less studied is the question whether or not these processes can be 
considered to be a DPT in the sense described above.
The main problem is similar to that appearing in the description of
the Dicke superradiance, an effect known for many years \cite{dicke}, 
however not fully understood up to today. In both cases, the experimental
studies are performed in optics. While the formal 
equivalence of the quantum mechanical Schr\"odinger equation and the 
optical wave equation in  symmetric optical lattices 
\cite{equivalence1,equivalence2,equivalence3,equivalence4}
is explored in the first case  for an
interpretation of the experimental results, a comparable theoretical 
study does not exist in the second case, i.e. for the Dicke superradiance.

It is the aim of the present paper to study the meaning of
the mathematical non-trivial properties of non-Hermitian operators
for the physics of open quantum systems that are embedded into a 
{\it common well-defined} environment. The mathematical properties are the
existence of singular points (EPs); the reduced phase rigidity ($r_i$) in
their vicinity; the appearance of nonlinear terms in the
Schr\"odinger equation due to  $r_i <1$; and the appearance of
{\it constructive} interferences. The physical observable
effects are DPTs known to appear at high level density. They will be
discussed in the following paper \cite{II}.

In our calculations we use a schematic model to simulate typical features 
of open quantum systems that are induced {\it coherently by the common
environment}. The obtained results are generic. The basic formalism 
used by us, is worked out in nuclear physics many years ago 
\cite{feshbach} where it is, however,  used by
introducing the non-Hermiticity  by means of a perturbation and, 
furthermore,  by using
statistical assumptions for the individual states (mostly
according to random matrix theory). In contrast to this, we consider 
directly the individual eigenvalues $\ce_i$ and eigenfunctions 
$\Phi_i$ of the non-Hermitian
Hamiltonian $\ch$.  In particular, we are interested in the influence
of the  EPs onto these values. As very well known, the
eigenvalues show level repulsion and (or) width bifurcation. We show
that the eigenfunctions contain new information, because they characterize 
the parameter range over which the influence of the EPs can be seen
and the manner how different EPs may influence each another. 

In the present paper, we consider a two-level system with real, 
complex and imaginary coupling coefficients between system and environment
with {\it loss} (emission) of particles to the environment (Sect. \ref{two1}) 
which is the usual situation of quantum systems embedded into the
environment of scattering wavefunctions. In Sect. \ref{two2}
we consider systems in which additionally {\it gain} (absorption) of particles 
from the environment occurs what is discussed recently in literature, e.g. 
\cite{joglekar,peng1,peng2,peng3}. 

In a following paper \cite{II}, we  address finally the problem of the 
relation between EPs and DPTs in systems with more than two nearby
states  coupled via a common environment. Here different EPs 
may influence each other.

\section{Crossing of two states in an open quantum system with symmetric 
non-Hermitian Hamiltonian}
\label{two1}

\subsection{Basic equations, Hamiltonian near an exceptional point}
\label{mod1}

In an open quantum system, the discrete states  described by a
Hermitian Hamiltonian $H^B$, are embedded into the continuum of scattering
wavefunctions, which exists always and can not be deleted. 
Due to this fact the discrete states turn into
resonance states the lifetime of which is usually finite. 
The Hamiltonian $\ch$ of the system which is embedded into the environment,
is non-Hermitian. Its eigenvalues are complex and
provide not only the energies  of the states but also their lifetimes
(being inverse proportional to the widths).
   
According to \cite{feshbach}, the  non-Hermitian Hamiltonian of an open 
quantum system  reads \cite{top} 
\begin{eqnarray}
\label{ham1}
\ch^F & = & H^B + V_{BC} G_C^{(+)} V_{CB} 
\end{eqnarray}
where the second term is the non-Hermitian perturbation;
$V_{BC}$ and $V_{CB}$ stand for the interaction between system
and environment; and $  G_C^{(+)} $ is the Green function in the
environment. The so-called internal (first-order) interaction 
between two discrete states $i$ and $j$ is
involved in $H^B$ while their external (second-order) interaction via the
common environment is described by the last term of (\ref{ham1}). 
Generally, the coupling matrix elements that determine the external
interaction of two states consist of the principal value integral    
\begin{eqnarray}
{\rm Re}\; 
\langle \Phi_i^{B} | \ch |  \Phi_j^{B} \rangle 
 -  E_i^B \delta_{ij} =\frac{1}{2\pi} 
 {\cal P} \int_{\epsilon_c}^{\epsilon_{c}'}  {\rm d} E' \;  
\frac{\gamma_{ic}^0 \gamma_{jc}^0}{E-E'} 
\label{form11}
\end{eqnarray}
which is real, and the residuum
\begin{eqnarray}
{\rm Im}\; \langle \Phi_i^{B} | \ch |
  \Phi_j^{B} \rangle =
- \frac{1}{2}\;  \gamma_{ic}^0 \gamma_{jc}^0 
\label{form12}
\end{eqnarray}
which is imaginary \cite{top}. Here, the $\Phi_i^{B}$ and  $E_i^B$ are the
eigenfunctions and (discrete) eigenvalues, respectively, of
the Hermitian Hamiltonian $H^B$ which describes the states in the subspace of
discrete states without any interaction of the states via the
environment. The $\gamma_{i c}^0  \equiv
\sqrt{2\pi}\, \langle \Phi_i^B| V | \xi^{E}_{c}\rangle $ 
are the (energy-dependent) coupling matrix elements  
between the discrete states $i$ of the system and the environment of
scattering wavefunctions $\xi_c^E$. The $\gamma_{k c}^0$ have to be
calculated for every state $i$ and for every channel $c$ 
(for details see \cite{top}). 
When $i=j$, (\ref{form11}) and (\ref{form12}) give the selfenergy of
the state $i$. 
The coupling matrix elements (\ref{form11}) and (\ref{form12})
(by adding $E_i^B \delta_{ij}$ in the first case) 
are often simulated by complex values  $\omega_{ij}$, e.g. 
\cite{elro2,comment2}. 

In order to study the interaction of two states via the 
common environment it is best to start from a 
non-Hermitian Hamiltonian $\ch$
in which $H^B$ in (\ref{ham1}) is replaced by a non-Hermitian 
Hamilton operator $\ch_0$ the eigenvalues of which are complex
(and not discrete as those of $H^B$). 
Let us consider, for example, the symmetric $2\times 2$ matrix 
\begin{eqnarray}
{\cal H}^{(2)} = 
\left( \begin{array}{cc}
\varepsilon_{1} \equiv e_1 + \frac{i}{2} \gamma_1  & ~~~~\omega_{12}   \\
\omega_{21} & ~~~~\varepsilon_{2} \equiv e_2 + \frac{i}{2} \gamma_2   \\
\end{array} \right) 
\label{form1}
\end{eqnarray}
with $\gamma_i \le 0$. The diagonal elements of (\ref{form1}) 
are the two complex eigenvalues 
$ \varepsilon_{i}~(i=1,2)$ of the non-Hermitian operator ${\cal H}_0$. That means,
the $e_i$ and  $\gamma_i$ denote the 
energies and widths, respectively, of the two states when
$\omega_{ij} =0$. The $\omega_{12}=\omega_{21}\equiv \omega$ stand for the coupling
matrix elements of the two states via the common environment which are, generally,
complex due to (\ref{form11}) and (\ref{form12}). The selfenergy of the
states is assumed to be included into the $\varepsilon_i$.
The Hamiltonian $\ch^{(2)}$ allows us to consider the properties of
  the system near to and at an EP because here the distance between
  the two states, that coalesce at the EP, relative to one another is
  much smaller than that relative to the other states of the system.
 Note that the coupling matrix elements  $\gamma_{k c}^0$ 
in (\ref{form11}) and (\ref{form12}) have the
dimension of square root of energy while the widths $\gamma_k$ of the
individual eigenstates in (\ref{form1})
have, of course, the dimension of energy.

\subsection{Eigenvalues of $\ch^{(2)}$}
\label{mod12}

The eigenvalues of $\ch^{(2)}$ are
\begin{eqnarray}
\ce_{i,j} \equiv E_{i,j} + \frac{i}{2} \Gamma_{i,j} = 
 \frac{\varepsilon_1 + \varepsilon_2}{2} \pm Z ~; \quad \quad
Z \equiv \frac{1}{2} \sqrt{(\varepsilon_1 - \varepsilon_2)^2 + 4 \omega^2}
\label{int6}
\end{eqnarray}
where  $E_i$ and $\Gamma_i$ stand for the
energy and width, respectively, of the eigenstate $i$. 
When the energy detuning of the two levels is varied, different
  behaviors of the eigenvalues (\ref{int6})
will be observed which depend on the coupling strength
$\omega$  between the states and their environment. Generally,
resonance states with nonvanishing widths $\Gamma_i$ 
repel each other in energy  according to  Re$(Z) $
while the widths bifurcate according to  Im$(Z)  $.
The transition from level repulsion to width bifurcation is studied 
numerically in e.g. \cite{elro3}.
The two states cross when $Z=0$. This crossing point is 
an EP according to the definition of Kato \cite{kato}.
Here, the two eigenvalues  coalesce, $\ce_{1}=\ce_{2}$.

According to (\ref{int6}), two interacting discrete states (with
$\gamma_1 = \gamma_2 =   0$ and $e_1 \ne e_2$) avoid always crossing since 
$\omega$ and $\varepsilon_1 - \varepsilon_2$ are real in this case and the 
condition $Z=0$ can not be fulfilled,
\begin{eqnarray}
(e_1 - e_2)^2 +4\, \omega^2 &>& 0 \; .  
\label{int6a}
\end{eqnarray}
In this case, the EP can be found only by
analytical continuation into the continuum. This situation is called usually
avoided crossing of discrete states.
It holds also for narrow resonance states if $Z=0$ cannot be
fulfilled due to the small widths of the two states. 
The physical meaning of this result is very well known since many
years: the avoided crossing of two discrete states at a certain critical
parameter value \cite{landau1,landau2} means that the
two states are exchanged at this  point, including their 
populations ({\it  population transfer}).  
 
When $\omega = i ~\omega_0 $ is imaginary, 
\begin{eqnarray}
Z = \frac{1}{2} \sqrt{(e_1-e_2)^2 - \frac{1}{4} (\gamma_1-\gamma_2)^2 
+i(e_1-e_2)(\gamma_1-\gamma_2) - 4\omega_0^2}
\label{int6i}
\end{eqnarray}
is complex. The condition  $Z= 0$ can be fulfilled only when $
(e_1-e_2)^2 - \frac{1}{4} (\gamma_1-\gamma_2)^2 =  4\omega_0^2$
and $(e_1-e_2)(\gamma_1-\gamma_2) =0$, i.e. when $\gamma_1 = \gamma_2$ 
(while $e_1 \ne e_2$). In this case 
\begin{eqnarray}
(e_1 - e_2)^2 -4\, \omega_0^2 &= &0 
~~\rightarrow ~~e_1 - e_2 =\pm \, 2\, \omega_0 \; ,
\label{int6b}
\end{eqnarray}
and two EPs appear.  It holds further
\begin{eqnarray}
\label{int6c}
(e_1 - e_2)^2 >4\, \omega_0^2 &\rightarrow& ~Z ~\in ~\Re \\
\label{int6d}
(e_1 - e_2)^2 <4\, \omega_0^2 &\rightarrow&  ~Z ~\in ~\Im 
\end{eqnarray}
independent of the parameter dependence of the $e_i$.
In the first case, the eigenvalues ${\cal E}_i = E_i+i/2\, \Gamma_i$ 
differ from the original values 
$\varepsilon_i = e_i + i/2~\gamma_i$ by a contribution to
the energies and in the second case by a contribution to the widths. 
The width bifurcation starts in the very neighborhood of one of the EPs and
becomes maximum in the middle between the two EPs.
This happens at the crossing point $e_1 = e_2$ where 
$\Delta \Gamma/2 \equiv |\Gamma_1/2 - \Gamma_2/2| = 4\, \omega_0$.
A similar situation appears when $\gamma_1 \approx \gamma_2$,  see
numerical results in Sect. \ref{simul1}.
The physical meaning of this result is completely different from that
discussed above for discrete and narrow resonance states.
It means that {\it different time scales} may appear without any
enhancement of the coupling strength to the continuum (for details see
\cite{fdp1}).

The cross section can be calculated by means of the $S$ matrix 
$\sigma (E) \propto |1-S(E)|^2$. For an isolated resonance, it gives 
the well-known Breit-Wigner line shape according to 
\begin{eqnarray}
\label{breitwigner1}
S=1+i\, \frac{\Gamma_1}{E-E_1 -\frac{i}{2}\Gamma_1}
\end{eqnarray}
where $E$ is the energy and $E_1$ and $\Gamma_1$ are defined in
  Eq. (\ref{int6}). 
This expression can be rewritten as \cite{ro03}  
\begin{eqnarray}
\label{breitwigner2}
S = \frac{E-E_1+\frac{i}{2}\Gamma_1}{E-E_1-
\frac{i}{2}\Gamma_1}
\end{eqnarray}
which is explicitly unitary  when the energy dependence of the
  $E_i$ and $\Gamma_i$ is taken into account \cite{top}.  
Extending the problem to that of two closely neighboring
resonance states that are coupled to one 
common continuum of scattering wavefunctions 
the unitary representation (\ref{breitwigner2})
of the $S$ matrix reads (up to a background term)  \cite{top} 
\begin{eqnarray}
\label{sm1}
S = \frac{(E-E_1+\frac{i}{2}\Gamma_1)~(E-E_2+\frac{i}{2}\Gamma_2)}{(E-E_1-
\frac{i}{2}\Gamma_1)~(E-E_2-\frac{i}{2}\Gamma_2)} \; .
\end{eqnarray}
In this expression, the influence of an EP onto the cross section 
is contained in the eigenvalues 
${\cal{E}}_i = E_i + i/2~\Gamma_i$ of $\ch^{(2)}$.
Reliable results can be obtained therefore  also when an EP is
approached and the $S$ matrix has a double pole at the 
parameter value corresponding to the EP. 
Here, the line shape of the two overlapping resonances is described by 
\begin{eqnarray}
\label{sm2}
S = 1+2i\frac{\Gamma_d}{E-E_d-\frac{i}{2}\Gamma_d}-
\frac{\Gamma_d^2}{(E-E_d-\frac{i}{2}\Gamma_d)^2}
\end{eqnarray}
by rewriting (\ref{sm1}),
where  $E_1=E_2\equiv E_d$  and $\Gamma_1=\Gamma_2\equiv \Gamma_d$.
It deviates  from the Breit-Wigner line shape of an isolated resonance 
due to  interferences between the two resonances. 
The first term of (\ref{sm2}) is
linear (with the factor $2$ in front) while the second one is  
quadratic. As a result, two peaks with asymmetric line shape
appear in the cross section (for a numerical example see Fig. 9 in 
\cite{mudiisro}).

\subsection{Eigenfunctions of $\ch^{(2)}$}
\label{mod13}

The eigenfunctions of a non-Hermitian $\ch$ 
must fulfill the conditions $\ch|\Phi_i\rangle =
  {\ce}_i|\Phi_i\rangle$ and $\langle \Psi_i|\ch = {\ce}_i \langle
  \Psi_i|$ where  $\ce_i$ is an eigenvalue of $\ch$ and the vectors 
$ |\Phi_i\rangle$ and $\langle \Psi_i|$ denote its right and left
eigenfunctions, respectively. When $\ch$ is a Hermitian operator, the 
 $\ce_i$ are real, and we arrive at the
 well-known relation $\langle \Psi_i| =  \langle \Phi_i|$. In this case, 
the eigenfunctions can be normalized by using the expression 
$\langle \Phi_i|\Phi_j\rangle$. 
For the symmetric non-Hermitian Hamiltonian $\ch^{(2)}$, however, we
have  $\langle \Psi_i| =  \langle \Phi_i^*|$. This means, the
eigenfunctions are biorthogonal and have to be normalized by means of 
$\langle \Phi_i^*|\Phi_j\rangle$. This is, generally, 
a {\it complex} value, in contrast to the real value 
 $\langle \Phi_i|\Phi_j\rangle$ of the Hermitian case. To smoothly
describe the transition from a closed system with discrete states, to
a weakly open one with narrow resonance states, we normalize the 
$\Phi_i$  according to 
\begin{eqnarray}
\langle \Phi_i^*|\Phi_j\rangle = \delta_{ij} 
\label{int3}
\end{eqnarray}
(for details see sections 2.2 and 2.3 of \cite{top}). It follows  
\begin{eqnarray}
 \langle\Phi_i|\Phi_i\rangle & = & 
{\rm Re}~(\langle\Phi_i|\Phi_i\rangle) ~; \quad
A_i \equiv \langle\Phi_i|\Phi_i\rangle \ge 1
\label{int4} 
\end{eqnarray}
and 
\begin{eqnarray}
\langle\Phi_i|\Phi_{j\ne i}\rangle & = &
i ~{\rm Im}~(\langle\Phi_i|\Phi_{j \ne i}\rangle) =
-\langle\Phi_{j \ne i}|\Phi_i\rangle 
\nonumber  \\
&& |B_i^j|  \equiv 
|\langle \Phi_i | \Phi_{j \ne i}| ~\ge ~0  \; .
\label{int5}
\end{eqnarray}
At an EP $A_i \to \infty$ and $|B_i^j| \to \infty$.
The $\Phi_i$ contain (like the $\ce_i$)  global features that are 
caused by many-body forces  induced by the coupling
$\omega_{ik}$ of the states $i$ and $k\ne i$ via the environment
(which has an infinite number of degrees of freedom).
The eigenvalues $\ce_i$ and eigenfunctions $\Phi_i$ contain 
moreover the self-energy contributions of the states $i$
due to their coupling to the environment. 

At the EP, the eigenfunctions $\Phi_i^{\rm cr}$ of ${\cal H}^{(2)}$
of the two crossing states differ from one another only by a phase, 
\begin{eqnarray}
\Phi_1^{\rm cr} \to ~\pm ~i~\Phi_2^{\rm cr} \; ;
\quad \qquad \Phi_2^{\rm cr} \to
~\mp ~i~\Phi_1^{\rm cr}   
\label{eif5}
\end{eqnarray}  
according to analytical  as well as numerical and experimental
studies, see  Appendix of \cite{fdp1}, section 2.5 of \cite{top}
and Figs. 4 and 5 in \cite{berggren}.
That means, the wavefunction $\Phi_1$ of the state $1$ jumps, 
at the EP, via the wavefunction ~$\Phi_1\pm \, i\, \Phi_2$  
of a chiral state to   ~$\pm\, i\, \Phi_2$
\cite{comment}. 

The Schr\"odinger equation with the non-Hermitian operator 
${\cal H}^{(2)}$ is equivalent to a Schr\"odinger equation with 
${\cal H}_0$ and source term \cite{ro01}
\begin{eqnarray}
\label{form1a}
({\cal H}_0 - \varepsilon_i) ~| \Phi_i \rangle  = -
\left(
\begin{array}{cc}
0 & \omega_{ij} \\
\omega_{ji} & 0
\end{array} \right) |\Phi_j \rangle \equiv W  |\Phi_j \rangle\; . 
\end{eqnarray}
Due to the source term, two states are coupled via the 
common environment of scattering wavefunctions into which the system 
is embedded,  $\omega_{ij}=\omega_{ji}\equiv\omega$.
The Schr\"odinger equation (\ref{form1a}) with source term can be
rewritten in the following manner \cite{ro01},
\begin{eqnarray}
\label{form2a}
({\cal H}_0  - \varepsilon_i) ~| \Phi_i \rangle  = 
\sum_{k=1,2} \langle
\Phi_k|W|\Phi_i\rangle \sum_{m=1,2} \langle \Phi_k |\Phi_m\rangle 
|\Phi_m\rangle \; . 
\end{eqnarray}
According to the biorthogonality  relations
(\ref{int4}) and (\ref{int5}) of the eigenfunctions of ${\cal H}^{(2)}$,  
(\ref{form2a}) is a nonlinear equation.  
Most important part of the nonlinear contributions is contained in 
\begin{eqnarray}
\label{form3a}
({\cal H}_0  - \varepsilon_n) ~| \Phi_n \rangle =
\langle \Phi_n|W|\Phi_n\rangle ~|\Phi_n|^2 ~|\Phi_n\rangle \; .  
\end{eqnarray}
The nonlinear source term vanishes far from an EP where
$\langle \Phi_k|\Phi_{k }\rangle    \to  1 $ and
$\langle \Phi_k|\Phi_{l\ne k }\rangle = - 
\langle \Phi_{l \ne k  }|\Phi_{k}\rangle \to  0 $
as follows from the normalization (\ref{int3}).
Thus, the Schr\"odinger equation with source term is 
linear far from an EP, as usually assumed. It is however nonlinear
in the neighborhood of an EP.

The biorthogonality of the eigenfunctions $\Phi_k$ of the non-Hermitian
operator $\ch^{(2)}$ is determined quantitatively by the ratio 
\begin{eqnarray}
r_k ~\equiv ~\frac{\langle \Phi_k^* | \Phi_k \rangle}{\langle \Phi_k 
| \Phi_k \rangle} ~= ~A_k^{-1} \; . 
\label{eif11}
\end{eqnarray}
Usually  $r_k \approx 1$ for decaying states which are well
separated  from other decaying states (according to the
fact that Hermitian quantum physics is a good approach at low level 
density). The situation changes however
completely when an EP is approached\,:
\begin{verse}
(i) When  two levels are distant from one another,  their eigenfunctions
 are (almost) orthogonal,  
$\langle \Phi_k^* | \Phi_k \rangle   \approx
\langle \Phi_k | \Phi_k \rangle  \equiv A_k \approx 1 $.\\
(ii) When  two levels cross at the EP, their eigenfunctions are linearly
dependent according to (\ref{eif5}) and 
$\langle \Phi_k | \Phi_k \rangle \equiv A_k \to \infty $.\\
\end{verse}
These two relations show that the phases of the two eigenfunctions
relative to one another change dramatically 
when the crossing point (EP) is approached. We call  $r_k$, 
defined by (\ref{eif11}), the {\it phase
  rigidity}  of the eigenfunction $\Phi_k$. Generally $1 ~\ge ~r_k ~\ge ~0 $.  
The  non-rigidity $r_k$ of the phases of the eigenfunctions of $\ch^{(2)}$ 
follows directly from the fact that $\langle\Phi_k^*|\Phi_k\rangle$
is a complex number (in difference to the norm
$\langle\Phi_k|\Phi_k\rangle$ which is a real number) 
such that the normalization condition
(\ref{int3}) can be fulfilled only by the additional postulation 
Im$\langle\Phi_k^*|\Phi_k\rangle =0$ (what corresponds to a rotation). 

When $r_k<1$, an analytical expression for the eigenfunctions as 
function of a certain control parameter  can, generally, not be
obtained. The  non-rigidity $r_k<1$ of the phases of the eigenfunctions
of $\ch^{(2)}$ in the neighborhood of EPs is the most important
difference between the  non-Hermitian quantum physics and the Hermitian
one. It expresses the fact that two nearby states can strongly 
interact with one another, when their wavefunctions are not supposed
to be everywhere orthogonal (as in Hermitian quantum physics). 
Mathematically, $r_k<1$ causes nonlinear effects in 
quantum systems in a natural manner, as shown above.
Physically, it gives the possibility that one of the  states of the system 
aligns at (or near to) the EP with the common environment and receives,
by this, a large width. This alignment is nothing but a
quantitative measure of the influence of the environment onto
the spectroscopic properties of the system \cite{top}. 

It is meaningful to represent
the  eigenfunctions $\Phi_i$ of ${\cal H}^{(2)}$  in the
set of basic wavefunctions $\Phi_i^0$ of ${\cal H}_0$
\begin{eqnarray}
\Phi_i=\sum_{j=1}^N b_{ij} \Phi_j^0 ~~ ;
\quad \quad b_{ij} = |b_{ij}| e^{i\theta_{ij}}
\; .
\label{int20}
\end{eqnarray}
Also the $b_{ij}$ are normalized  according to the biorthogonality
relations  of the wavefunctions $\{\Phi_i\}$. The angle $\theta_{ij}$
can be determined from
${\rm tg}(\theta_{ij}) = {\rm Im}(b_{ij}) / {\rm Re}(b_{ij})$ .

It should be mentioned here that the eigenfunctions  $\Phi_k$ of 
${\cal H}^{(2)}$ represent only the  part of the resonance wavefunction
that is localized inside the system. The  wavefunction
of the resonance state $k$ in the  whole function
space of discrete and scattering states contains additionally a ``tail'' 
due to its coupling to the scattering wavefunctions, see \cite{top}.

\subsection{Numerical results}
\label{simul1}

\begin{figure}[ht]
\begin{center}
\includegraphics[width=12cm,height=12cm]{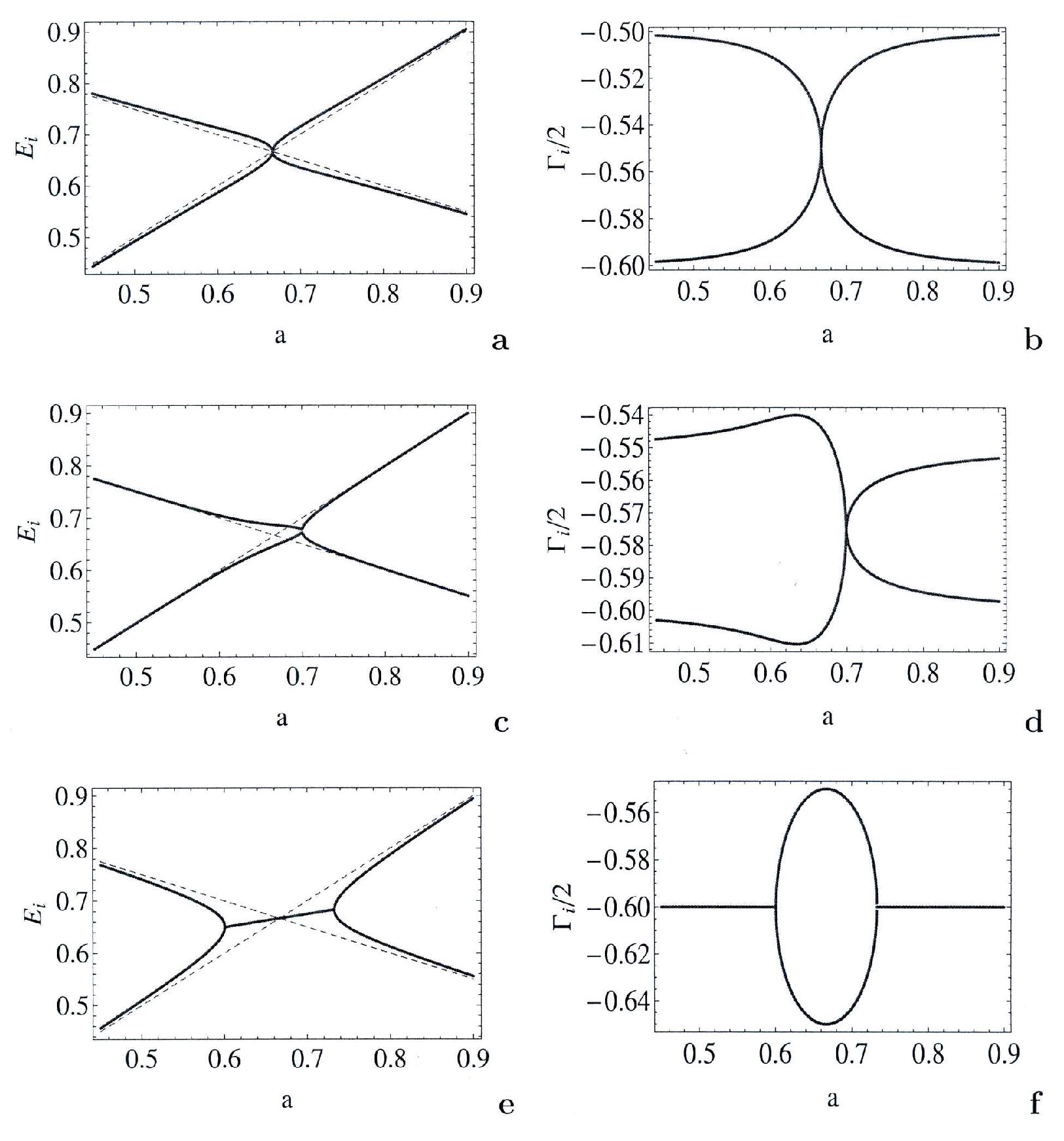}
\vspace{-.6cm}
\end{center}
\caption{\footnotesize
Energies $E_i$ (left panel) and widths $\Gamma_i/2$ (right panel) of 
$N=2$ states coupled to a
common channel as a function of $a$. Parameters: $e_1=1-0.5~a; ~~e_2=a$;
 ~~$\gamma_1/2= - 0.5$ (a,b);  ~ --0.5505 (c,d);  ~ --0.6 (e,f); 
~~$\gamma_2/2= - 0.6$; 
~~$\omega = 0.05$ (a,b); ~0.025~(1+i) (c,d); ~$0.05~i$ (e,f).
The dashed lines in (a, c, e) show $e_i(a)$.
}
\label{fig1}
\end{figure}

\begin{figure}[ht]
\begin{center}
\includegraphics[width=12cm,height=12cm]{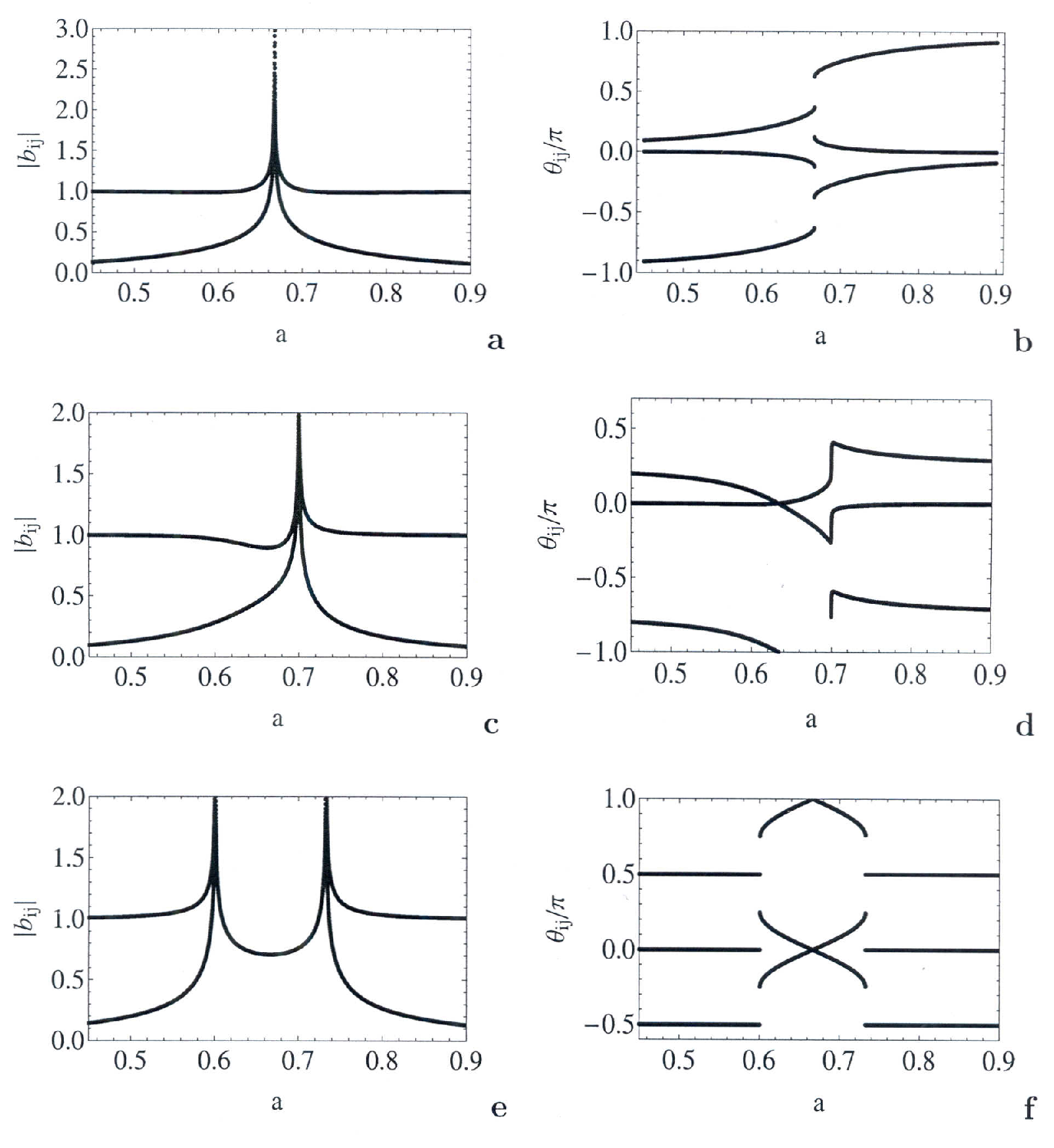}
\vspace{-.6cm}
\end{center}
\caption{\footnotesize
Mixing coefficients $b_{ij}= |b_{ij}|e^{i\theta_{ij}}$ of $N=2$ states coupled to
a common channel as a function of $a$. The parameters are the same as in
Fig. \ref{fig1}.
}
\label{fig2}
\end{figure}

In our calculations, the mixing coefficients $b_{ij}$, 
defined in (\ref{int20}), of the wavefunctions of the two  states  
are calculated by taking into account the fact  that the mixing  
depends on the distance (in energy) of the two states, what
can be simulated by assuming a  Gaussian distribution 
\begin{eqnarray}
\omega_{i\ne j} = \omega ~e^{-(e_i -e_j)^2}
\label{int7}
\end{eqnarray}
for the coupling coefficients. The results reproduce very well 
\cite{elro2,comment2} those
obtained numerically exact in \cite{ro01} for two levels and real 
coupling $\omega$. Further, the selfenergies of the states are
assumed, in our calculations, to be included into the $\varepsilon_i$.

Let us first consider the $2\times 2$ matrix (\ref{form1})
with $e_1=1-\frac{a}{2}; ~e_2=a $ and  with $\gamma_i ~(i=1,2)$ 
and $\omega_{12} = \omega_{21} \equiv \omega $ independent of $a$. 
For illustration, we show in Fig. \ref{fig1} the eigenvalue 
trajectories $E_{i}(a)$ and $\Gamma_{i}/2(a)$ and in Fig. \ref{fig2}
the mixing coefficients $b_{ij} = |b_{ij}|e^{i\theta_{ij}}$ 
(defined in (\ref{int20})) of 
the eigenfunctions of $\ch^{(2)}$ as a function of $a$ in the 
neighborhood of an EP. The calculations are performed with real,
complex and imaginary coupling coefficients $\omega$. Both, the
upper (real $\omega$) and middle (complex $\omega$) rows of Figs. 
\ref{fig1} and \ref{fig2} show an EP at the 
critical parameter value $a=a_{\rm cr}$. Here the eigenvalue
trajectories cross and  $|b_{ij}| \to \infty$. 
The lower row is calculated with imaginary $\omega$ and $\gamma_1 =
\gamma_2$. Here two EPs appear, and $|b_{ij}| \to \infty$ at every EP. 

The main difference of the eigenvalue trajectories with
real to those with imaginary coupling coefficients $\omega$ are
related to the relations (\ref{int6a}) to (\ref{int6d}) obtained
analytically and discussed in Sect. \ref{mod12}.
For real and complex $\omega$ and $\gamma_1 \ne \gamma_2$, the
results show one EP (when the condition $Z=0$
is fulfilled, see Fig. \ref{fig1}, upper and middle rows).  
This EP is isolated from other EPs, generally. In the case of
imaginary $\omega$ and $\gamma_1 \approx \gamma_2$,
however, two related EPs appear (Fig. \ref{fig1}, lower row).
Between these two EPs, the widths $\Gamma_i$ bifurcate: the width of
one of the two states increases  by varying $a$ although  
the coupling strength $\omega$ between system and environment
remains constant. 

As can be seen from Fig. \ref{fig2} left panel, the critical parameter 
range has a finite extension at both sides of the EPs. 
When $\omega$ is imaginary, the critical parameter range
includes both EPs and their vicinity. Between the two EPs the 
eigenfunctions are strongly mixed (1:1) with one another. 
Beyond the critical parameter region, 
the eigenvalues trajectories $\ce_{i}(a)$ approach
the trajectories $\varepsilon_j(a)$ after exchange of $i$ and $j$.

Interesting are also the phases of the eigenfunctions in the 
neighborhood of an EP, see Fig. \ref{fig2} right panel. 
The phases of {\it all} components of the
eigenfunctions jump at the EP either by $-\pi /4$ or by
$+\pi /4$. That means the phases of {\it both} eigenfunctions
jump in the same direction by the same amount. Thus, there is a phase
jump of $-\pi /2$ (or $+\pi /2$) when one  of the eigenfunctions passes into   
the other one at the EP. This result is in agreement with (\ref{eif5}).
It holds true for real as well as for complex and imaginary  
$\omega$ as can be seen from Fig. \ref{fig2} right panel. 

The position of an isolated EP can always be found by 
varying another parameter. For example, with 
$e_1=1-\frac{a}{2}+r\, {\rm cos}\theta; ~e_2=a+r\, {\rm sin}\theta$
one EP appears in any case in the parameter range $0 \le \theta \le \pi$. The
results obtained in the neighborhood of and at this  EP  show the same 
characteristic features as those in Figs. \ref{fig1} and \ref{fig2}: 
around the crossing
point (EP) of the eigenvalue trajectories, the eigenfunctions 
are mixed and  $|b_{ij}|\to \infty$ at the EP. The phase jumps are of the
same type as those shown in Fig. \ref{fig2}, 
confirming the relation (\ref{eif5})
between the two eigenfunctions at the EP also by these calculations.

\begin{figure}[ht]
\begin{center}
\includegraphics[width=13cm,height=9cm]{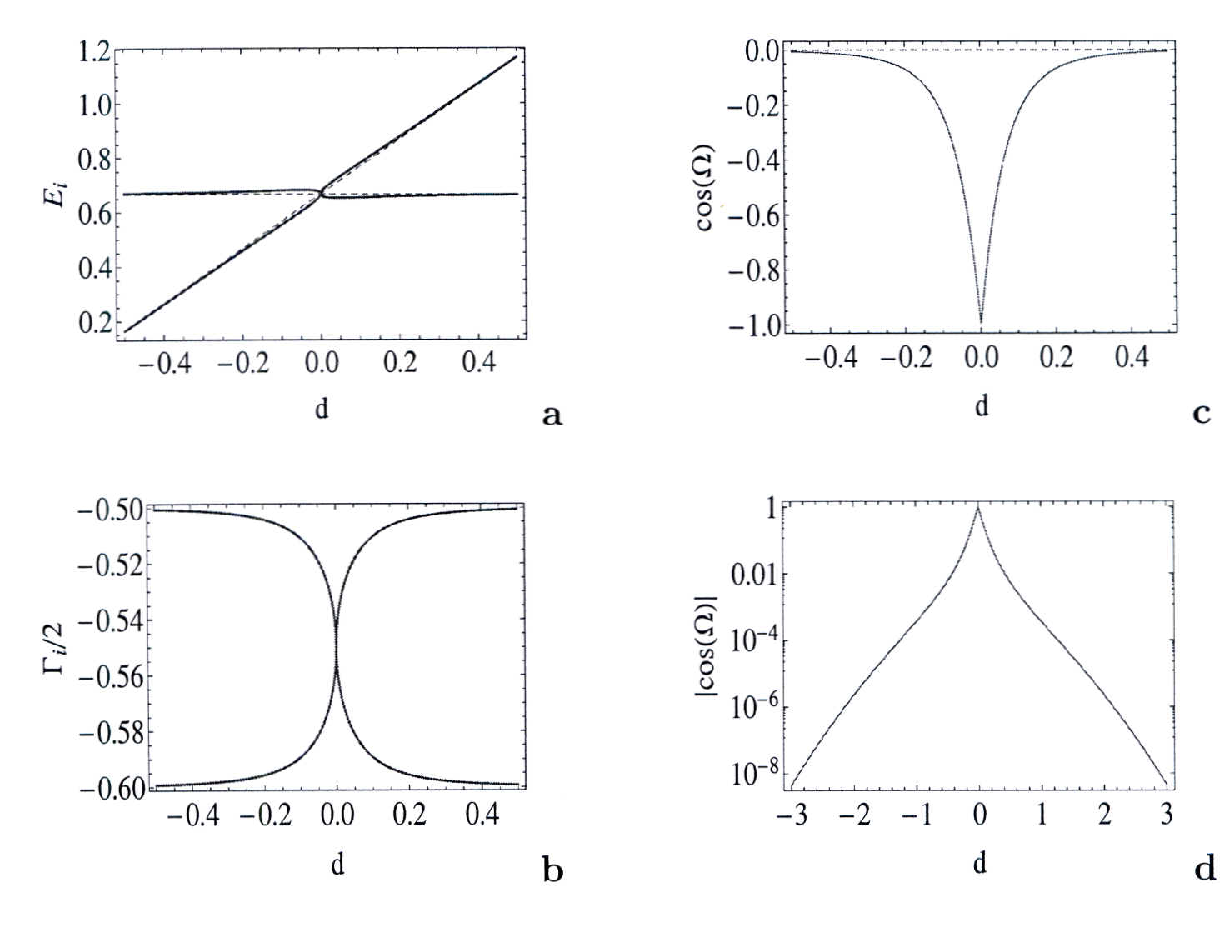}
\vspace{-.6cm}
\end{center}
\caption{\footnotesize
{Energies $E_{i}$ (full lines) (a), widths $\Gamma_{i}/2$ (b),
and cos$(\protect\Omega)$, (c) in linear scale, (d) in logscale, 
as function of
the distance $d$ for $N=2$ states coupled to one channel. The
unperturbed energies are $~e_{1}=2/3$ and $~e_{2}=2/3+d$ (dashed 
lines in (a)). The other parameters are $\protect\omega=0.05$, 
~$\protect\gamma_{1}/2=-0.5$, ~$\protect\gamma_{2}/2=-0.5999$. 
The dashed lines in (c, d) show
$|$cos$(\protect\Omega)|$ for the two orthogonal states of the Hermitian
operator $H^B$.}}
\label{fig1a}
\end{figure}

\begin{figure}[ht]
\begin{center}
\includegraphics[width=13cm,height=13cm]{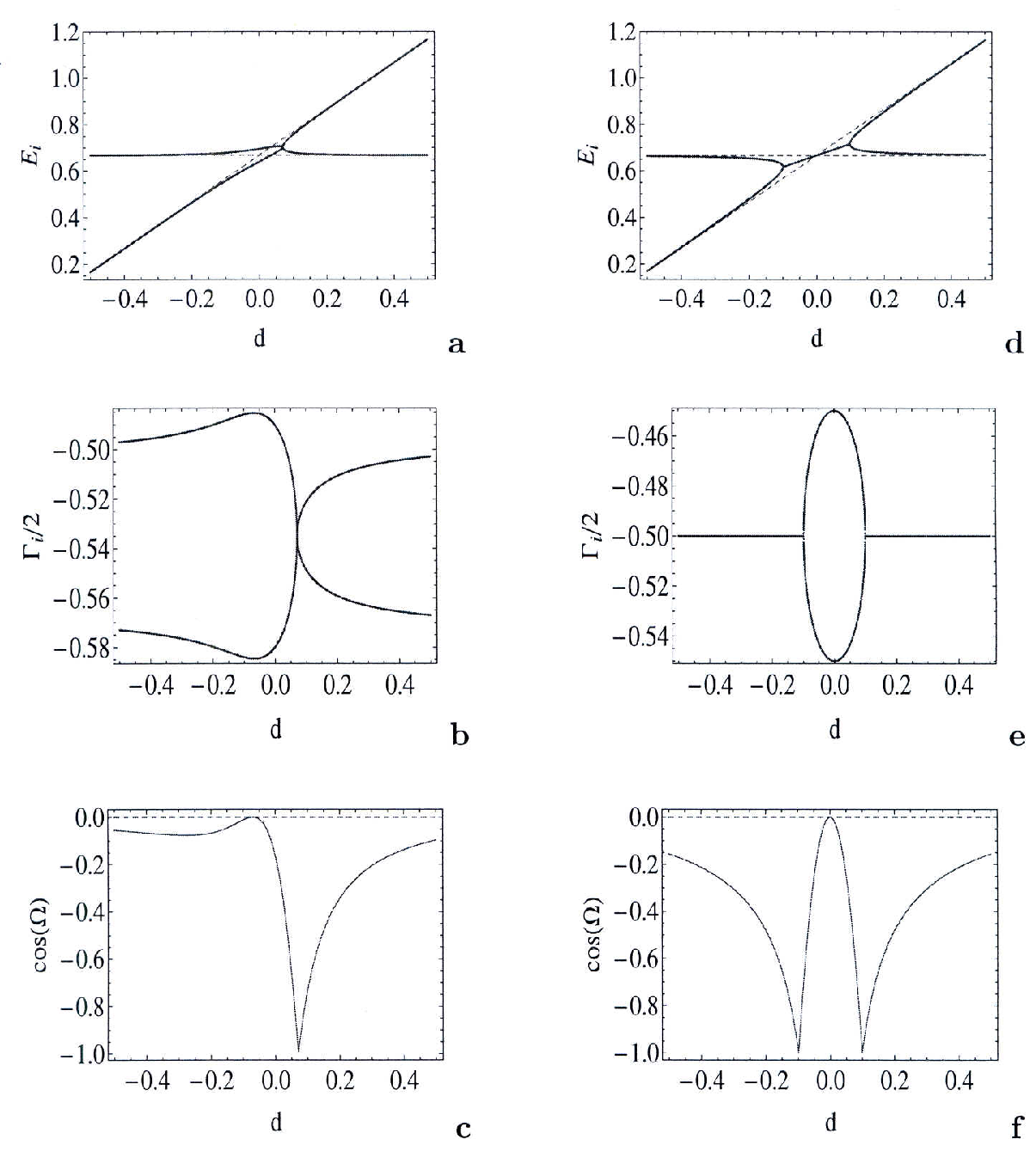} 
\vspace{-.6cm}
\end{center}
\caption{\footnotesize
{The same as Fig. \protect\ref{fig1a}.a-c, but $\protect\omega
=0.05 \, (1+i)/\protect\sqrt{2} $, ~$\protect\gamma_1/2=-0.5, 
~\protect\gamma_2/2=-0.57$ (left panel, a-c) and $\omega = 0.05\, i,
~\gamma_1/2=-0.5,  ~\gamma_2/2=-0.5$ (right panel, d-f).}
}
\label{fig1b}
\end{figure}

Now we explore numerically 
the phase difference $\Omega $ between the two eigenfunctions of the
operator $\mathcal{H}^{(2)}$ that describes a 2-level open quantum system.
The calculations are performed by starting from the unperturbed energies $
\varepsilon _{k}=e_{k}+\frac{1}{2}\gamma _{k}$ (diagonal matrix
elements of (\ref{form1})), with the assumptions that $e_{1}=\mathrm{const}$ while $
e_{2}=e_{2}(d)$ depends on the distance $d$ between the two states which
cross at $d=0$. The widths of both states are assumed to be constant, $
\gamma _{k}=$ const for $k=1,~2$. 
The angle $\Omega $ between the two eigenvectors of $
\mathcal{H}^{(2)}$ is represented in the figures by cos$(\Omega)$ in order
to illustrate the changes of $\Omega$ in approaching an EP. 
The coupling strength $\omega$  is chosen
to be real (Fig. \ref{fig1a}), complex (Fig. \ref{fig1b} left panel), and 
imaginary (Fig. 
\ref{fig1b} right panel). In the first and second case, we have one EP 
while in the last case, there are two EPs according to (\ref{int6b}).

The results shown in Figs. \ref{fig1a} and \ref{fig1b} are the
following. (i) For distant levels, the two 
eigenfunctions are almost orthogonal. Here, asymptotically  
cos$(\Omega) \approx 0$, however the value cos$(\Omega)$ never vanishes
(see Fig. \ref{fig1a}.d in logarithmic scale).
(ii) At the EP, the eigenfunctions are
linearly dependent from one another according to (\ref{eif5}), 
what is expressed by cos$(\Omega) \to \pm 1$ in approaching the EP. These
results confirm the statements according to which the
normalization of the eigenfunctions of a non-Hermitian operator by means of
the complex value (\ref{int3}) is possible only by rotating the eigenvector
such that Im$\langle\Phi_k^*|\Phi_k\rangle =0$. The rotation angle,
represented by cos$(\Omega)$, is shown in Figs. \ref{fig1a}.c,  and 
\ref{fig1b}.c and f for different values of the coupling coefficient $\omega$.

As can be seen from the eigenvalue equations (\ref{int6}) and from Figs.
\ref{fig1} to \ref{fig1b}, two states may
avoid crossing at the EP by level repulsion (as very well known since many
years \cite{landau1,landau2}), or they may cross freely while their widths bifurcate.
In the last case, the lifetimes of the two states may finally differ
strongly from one another, even \textit{bound states in the continuum} may
arise. The existence of these states is discussed already in the very early
days of quantum mechanics \cite{wigner}, later considered in atomic physics
\cite{wintgen1,wintgen2} and other systems, e.g. \cite{rosabic}. The eigenvalues of
the Hamiltonian show the existence and position of the critical parameter
values (corresponding to the EPs) at which level repulsion or width
bifurcation takes place.

Figs. \ref{fig1} to \ref{fig1b} illustrate furthermore how  an EP  influences 
its neighborhood and determines the dynamics of an open quantum system.
(i) The wavefunctions of the two crossing states are mixed and
the phases of the wavefunctions of the two states relative to one another 
vary in a {\it finite} parameter range in the neighborhood of the EP.
The reduction of the phase rigidity $r_k$  (corresponding to 
(\ref{eif11})) allows one of the states to {\it  align to the
states of the environment}, i.e. to receive a large width, while 
the other state almost decouples from the environment. 
(ii) When the interaction of the two states via the environment is
imaginary and the widths of both states are similar to one another
($\gamma_1 \approx \gamma_2$), width bifurcation occurs {\it between
the two EPs} according to (\ref{int6b}) and (\ref{int6d})
{\it without} any enhancement of the coupling strength to the environment.
The phases jump at the two EPs in different directions and
the eigenvalues approach the original values only beyond the two EPs. 

Figs. \ref{fig1} to \ref{fig1b} illustrate the most important
difference between Hermitian and non-Hermitian quantum physics\,: the phases
of the eigenfunctions of a Hermitian operator relative to one another are
fixed by the orthogonality relations at all parameter values, while those of
$\mathcal{H}^{(2)}$ are not everywhere rigid. They are influenced by the
singular points (EPs) at which two eigenvalues of the non-Hermitian operator
$\mathcal{H}$ coalesce. Here, the two eigenstates are exchanged, what is
accompanied by a change of the angle between the two eigenvectors according
to (\ref{eif5}). This process occurs not only at the position of the EP but
is characteristic for a certain finite parameter range around it, as can be
seen from the numerical results for the phase rigidity $r_k$ and for the
angle $\Omega$ between the two eigenvectors.

Of prime importance for physical processes induced by an EP in an open
quantum system that is embedded into a common well-defined
environment, are the nonlinear terms occurring in the Schr\"odinger
equation in the whole function space where $r_k <1$, see Eqs. (\ref{form2a})
and (\ref{form3a}). Eventually, they allow for some stabilization of the system 
by putting information on the environment into the system with the aim to
accumulate as much as possible of the total coupling strength between system
and environment onto one of the states (in the one-channel case). By this,
this state becomes short-lived while the other one decouples more or less
from the environment and becomes long-lived. These two states are not
analytically connected to the original individual states of the system. 

While the mathematical properties of the eigenvalues of ${\cal  H}^{(2)}$ 
are studied in many papers for isolated EPs,
their influence onto the vicinity of the EPs and onto the eigenfunctions 
is considered in only a few papers, see e.g. the review \cite{top}. 
The interesting question how the ranges of different EPs may influence
each other is not at all considered in the literature. It will be  
discussed in detail in the following paper \cite{II} by using the results
shown in Figs. \ref{fig1} to \ref{fig1b}.

\section{Crossing of two states in quantum systems with loss and gain}

\label{two2}

\subsection{Basic equations,  Hamiltonian with loss and gain}
\label{mod2}

As has been shown in
\cite{equivalence1,equivalence2,equivalence3,equivalence4},
the quantum mechanical
Schr\"odinger equation and the optical wave equation 
in  symmetric optical lattices are formally equivalent. 
Complex symmetric structures can be realized by involving 
symmetric index guiding and an antisymmetric gain/loss profile. 

The main difference of
these optical systems to open quantum systems consists in the symmetry of
gain and loss in the first case while the states 
of an open quantum system can only decay (Im$(\varepsilon_{1,2}) < 0$
and Im$({\cal E}_{1,2}) < 0$
for both states).  Thus, the  modes involved in the non-Hermitian 
Hamiltonian in optics appear in complex conjugate pairs 
while this is not the case in an open quantum system.
As a consequence, the Hamiltonian for the description of the structures in 
optical lattices may have real eigenvalues in a large parameter range
\cite{elro4a}, similar as in, e.g., the papers 
\cite{bender,mumbai1,mumbai2,mumbai3}.

The $2\times 2$ non-Hermitian Hamiltonian may be written, in this case, as
\cite{opt,mumbai1,mumbai2} 
\begin{eqnarray}
\label{ham2o1}
\ch_{PT} =\left(
\begin{array}{cc}
~e-i\frac{\gamma}{2}~ &  w \\ w & ~e+i\frac{\gamma}{2}~
\end{array}
\right), 
\end{eqnarray}
where $ e $ stands for the energy  of the two modes, 
$\pm \gamma$ describes  gain and loss, respectively,
and the real coupling coefficient $w $ stands for the
coupling of the two modes via the lattice. 
When  optical lattices are studied with vanishing 
gain, the Hamiltonian reads 
\begin{eqnarray}
\label{ham2o2}
\ch_{PT}' =\left(
\begin{array}{cc}
~~e-i\frac{\gamma}{2}~~ &  w \\ w & e
\end{array}
\right) \; .
\end{eqnarray}
In realistic systems, $ w$ in (\ref{ham2o1}) and (\ref{ham2o2}) is
mostly  real (or at least almost real).

\subsection{Eigenvalues of the  Hamiltonian with loss and gain}
\label{mod22}

The eigenvalues of the Hamiltonian (\ref{ham2o1}) differ from 
(\ref{int6}),
\begin{eqnarray}
\label{eig2o1}
\ce^{PT}_\pm &=& e \pm \frac{1}{2}\sqrt{4|w|^2 - \gamma^2} \equiv e
\pm  Z_{PT} \, .
\end{eqnarray}
A similar expression is derived in \cite{mumbai1,mumbai2}.
Since $e$ and $\gamma$ are real, the $\ce^{PT}_\pm$ are real when
$4|w|^2 > \gamma^2$. Under this condition,
the two levels repel each other in energy what is
characteristic of discrete interacting states.
When the interaction $w$ is fixed, the level repulsion
decreases with increasing $\gamma$. When $4|w|^2 = \gamma^2 $  
the two states cross. Here,  $\ce^{PT}_\pm = e$ and $\gamma = \pm \sqrt{4|w|^2}$.
With further increasing $\gamma$ and $4|w|^2 < \gamma^2$
($w$ fixed for illustration), width bifurcation (called PT-symmetry breaking) 
occurs and $\ce^{PT}_\pm = e \pm \frac{i}{2}\sqrt{\gamma^2 - 4|w|^2}$.
 
These relations are in
accordance with (\ref{int6a})  to (\ref{int6d}) for open quantum systems.
Since $|w|$ is real,  two EPs exist according to  
\begin{eqnarray} 
\label{pt2}
4 |w|^2=(\pm \gamma)^2 \; . 
\end{eqnarray}
Further
\begin{eqnarray}
\label{pt3}
\gamma^2 <4\, |w|^2 &\rightarrow& ~Z_{PT} ~\in ~\Re \\
\label{pt4}
\gamma^2 >4\, |w|^2 &\rightarrow&  ~Z_{PT} ~\in ~\Im 
\end{eqnarray}
independent of the parameter dependence of $\gamma$.

In the case of the Hamiltonian (\ref{ham2o2}), the eigenvalues read
\begin{eqnarray}
\label{eig2o2}
\ce^{'PT}_\pm = e - i~\frac{\gamma}{4}
\pm \frac{1}{2}\sqrt{4|w|^2 - \frac{\gamma^2}{4}} \equiv e -
i~\frac{\gamma}{4} \pm Z_{PT}'.
\end{eqnarray}
We have level repulsion as long as 
$4|w|^2 > \frac{\gamma^2}{4}$. While  level repulsion 
decreases with increasing $\gamma$,
the loss increases with increasing $\gamma$.
At the crossing point, $\ce^{'PT}_\pm = e - i~\frac{\gamma}{4}$. 
With further increasing $\gamma$ and $ 4|w|^2 \ll \frac{\gamma^2}{4}$ 
\begin{eqnarray}
\label{eig2o3}
\ce^{'PT}_\pm \to  e - i~\frac{\gamma}{4} \pm i~\frac{\gamma}{4}
= \Bigg\{    \begin{array}{c}
e \qquad \\
e - i~\frac{\gamma}{2}.
\end{array} 
\end{eqnarray}
The two modes (\ref{eig2o3}) behave differently. While  loss in one
of them is large, it is almost zero in the other one. Thus, only one
of the modes effectively survives. Equation (\ref{eig2o3})
corresponds to high transparency at large $\gamma$.

Further, two EPs exist according to 
\begin{eqnarray} 
\label{pt6}
4 |w|^2 =(\pm \gamma /2)^2 
\end{eqnarray}
and
\begin{eqnarray}
\label{pt7}
\gamma^2/4 <4\, |w|^2 &\rightarrow& ~Z_{PT}^{'} ~\in ~\Re \\
\label{pt8}
\gamma^2/4 >4\, |w|^2 &\rightarrow&  ~Z_{PT}^{'} ~\in ~\Im \; . 
\end{eqnarray}
In analogy to (\ref{pt2}) up to (\ref{pt4}) these relations 
are independent of the parameter dependence of $\gamma$.

Thus, there exist similarities between the eigenvalues 
$\ce_i$ of $\ch^{(2)}$ of an  
open quantum system and the eigenvalues of the Hamiltonian of a 
system with gain and loss.
Interesting is the comparison of the eigenvalues $\ce_i$
of $\ch^{(2)}$ obtained for
imaginary non-diagonal matrix elements $\omega$, with the 
eigenvalues of (\ref{ham2o1}) or (\ref{ham2o2}) for real $w$.
In both cases, there are two EPs. In the first case,
the energies $E_i$ are constant and the widths $\Gamma_i$
bifurcate between the two EPs. This situation is characteristic of an
open quantum system at high level density with complex (almost
imaginary) $\omega$, see Eqs. (\ref{int6b}) to (\ref{int6d}).
In the second case however 
the difference $|E_1 - E_2|$ in the energies increases (level repulsion)
while the  widths $\Gamma_i$ of both states are equal in
the parameter range between the two EPs, see (\ref{pt2}) 
to (\ref{pt4}) and  (\ref{pt6}) to (\ref{pt8}), respectively.
Between the two EPs, level repulsion causes the
two levels to be distant from one to another and $w$ is expected to be
(almost) real according to (\ref{form11}) and (\ref{form12}). 
Formally, the role of energy and width is exchanged in the two cases. 

It should be underlined here that the non-Hermitian Hamiltonian
describing  an open quantum system may also have real eigenvalues if
certain conditions are fulfilled. Such a case is studied already more
than 80 years ago \cite{wigner},  later in atomic physics
\cite{wintgen1,wintgen2,marost1,marost2} and also in other systems 
such as double
quantum dots \cite{rosabic,top}. The so-called bound states in the
continuum are caused by width bifurcation and, consequently,
the width of the long-lived  
resonance state may approach zero. This mechanism is 
different from that considered here since it creates real eigenvalues of
the non-Hermitian Hamiltonian only at a few special parameter values.

\subsection{Eigenfunctions of the Hamiltonian with loss and gain}
\label{mod23}

The eigenfunctions of the two $2\times 2$ Hamiltonians (\ref{ham2o1})
and (\ref{ham2o2}) show the same characteristic features as those of 
the Hamiltonian (\ref{form1}).
The eigenmodes  can be normalized, generally, according to  
(\ref{int3}) where $ \Phi_i^{PT} ~(\Phi_i^{'PT})$ denotes the right eigenmode.
Far from an EP, the eigenfunctions $\Phi_i^{PT} ~(\Phi_i^{'PT})$ 
are orthogonal to one another. The
orthogonality is lost in approaching the crossing point of the
eigenvalue trajectories.
Here,  the modes show some skewness according to (\ref{int4}).
As in the case of open quantum systems,
the phase rigidity $r_i$ can be defined according to
(\ref{eif11}). It varies between 1 
and 0 and is a quantitative measure for the skewness of the modes. 
Thus, the phases of the eigenmodes of the
non-Hermitian Hamiltonians (\ref{ham2o1}) and (\ref{ham2o2})
are not  rigid, and  spectroscopic
redistribution processes may occur  
under the influence of the environment (lattice). 

The eigenfunctions $\Phi_i^{PT}$ of $\ch_{PT}$
(and $\Phi_i^{'PT}$ of $\ch_{PT}'$) can be represented in a
set of basic wavefunctions in full analogy to the representation of
the eigenfunctions $\Phi_i$ of $\ch^{(2)}$ in (\ref{int20}).
They contain valuable information on the mixing of the wavefunctions
under the influence of the non-diagonal coupling matrix elements $w$
in (\ref{ham2o1}) and (\ref{ham2o2}), respectively, 
as well as its relation to EPs.

\subsection{Numerical results for a quantum  system with loss and gain}
\label{simul2}

In realistic systems, the non-diagonal matrix elements $w$ of 
the non-Hermitian Hamiltonians (\ref{ham2o1}) and (\ref{ham2o2}) 
are real (or almost real)  as follows  from
the level repulsion occurring between the two EPs (see above, 
Sect. \ref{mod22}). Nevertheless, we did some calculations also for complex
and imaginary $w$ (results are not shown).

According to  (\ref{ham2o1}) and (\ref{ham2o2}), the energies $e_i$ 
and widths $\gamma_i$ of the two states are the same. We choose $e_1 =
e_2 \equiv e$
independent of the parameter $a$ in the considered region and 
$\gamma_i$ (gain and loss) to be parameter dependent.

\begin{figure}[ht]
\begin{center}
\includegraphics[width=12cm,height=12cm]{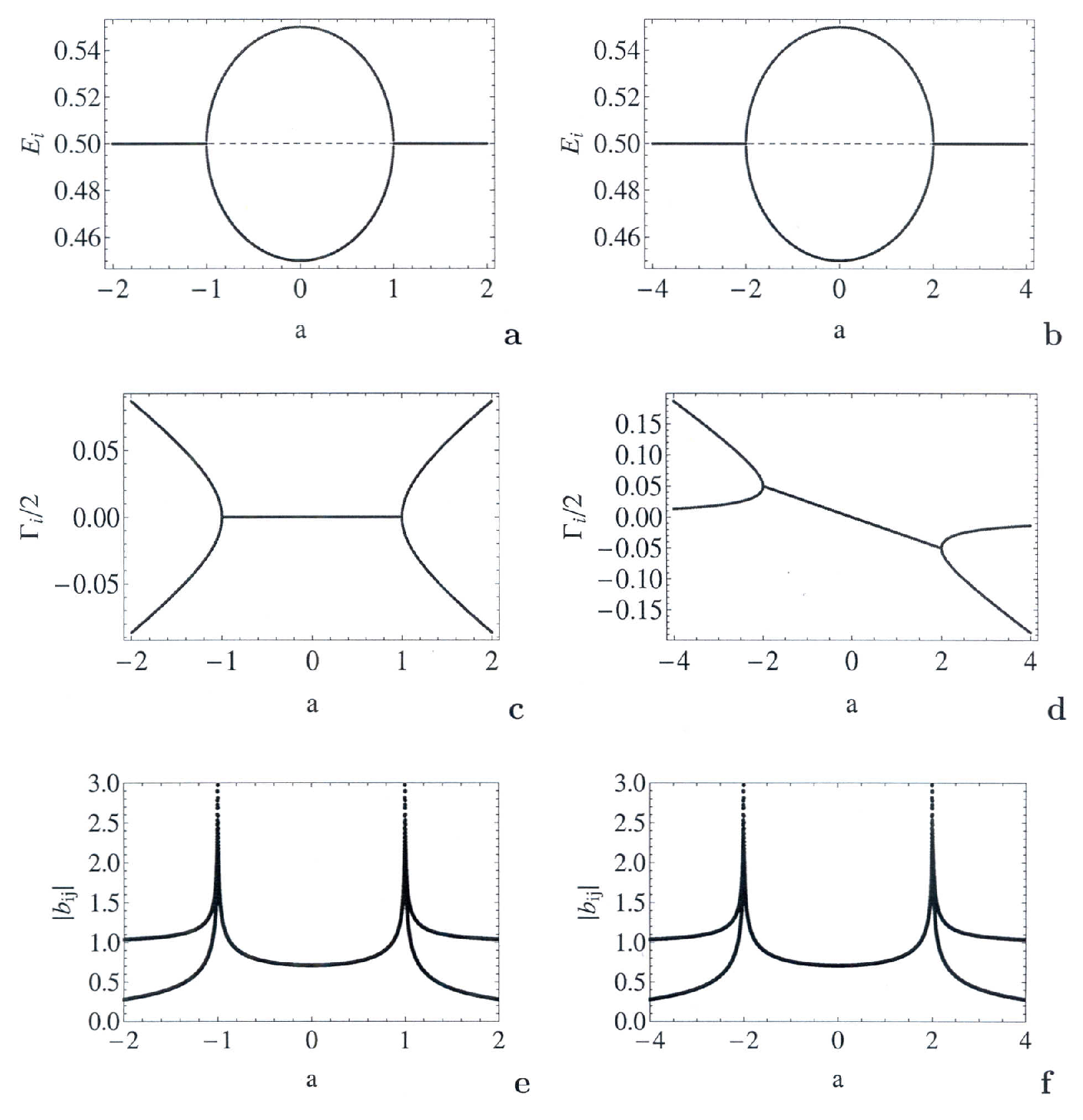}
\vspace{-.6cm}
\end{center}
\caption{\footnotesize
Energies $E_i$ (top), widths $\Gamma_i/2$ (mid) and mixing
coefficients $|b_{ij}|$ (bottom) of the eigenfunctions $\Phi_i$
of $N=2$ states coupled to a common channel as a function of
$a$. Parameters: 
$e=0.5; ~~w=0.05$;
~~$\gamma_1/2=- 0.05~a$; and $\gamma_2=-\gamma_1$ ~(left panel);
~~$\gamma_2=0$ ~(right panel).  
In order to illustrate the symmetry properties, the results are shown
for positive as well as for negative values $a$.
The dashed lines in (a, b) show $e$.
}
\label{fig3}
\end{figure}

\begin{figure}[ht]
\begin{center}
\includegraphics[width=12cm,height=12cm]{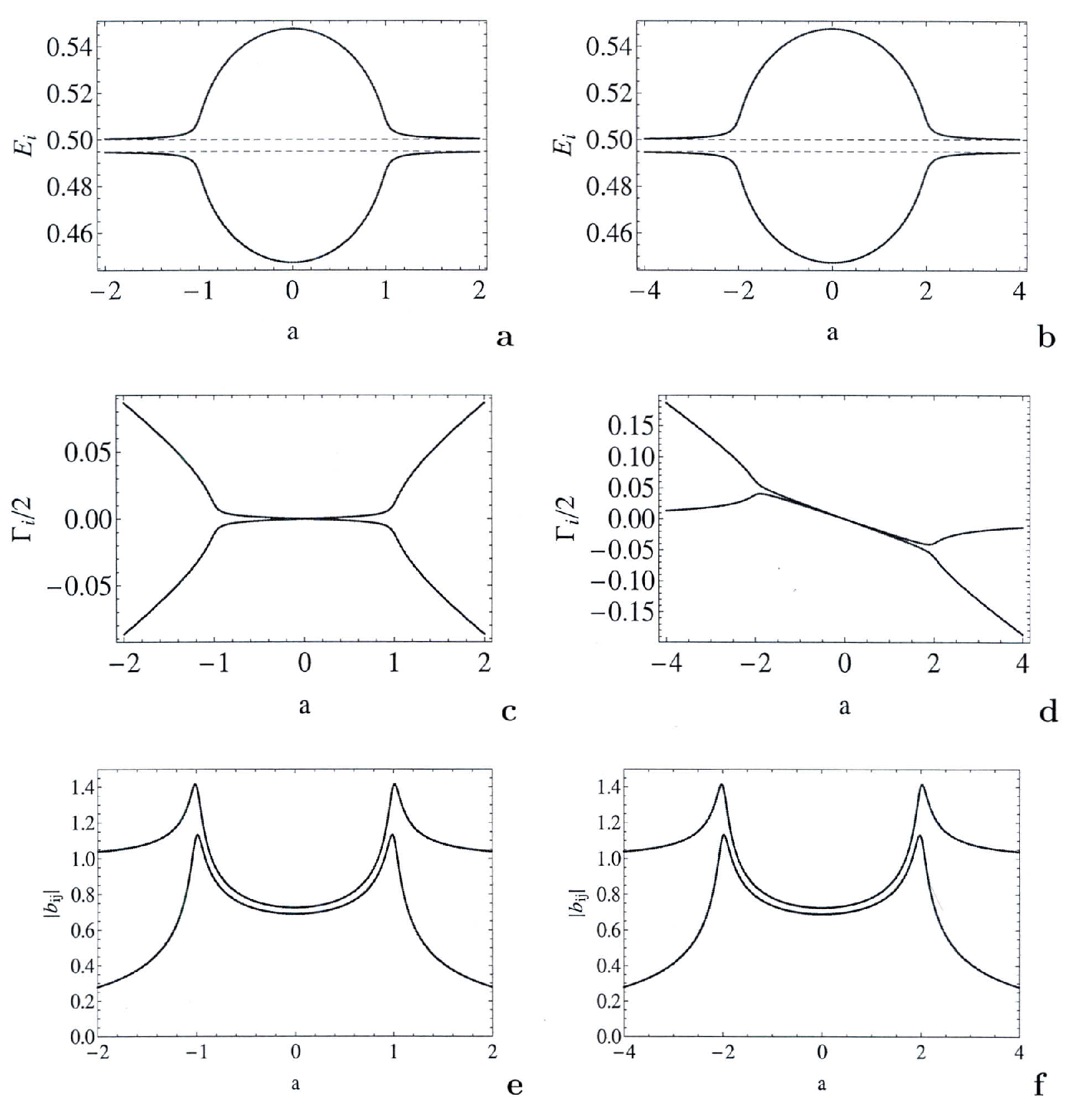}
\vspace{-.6cm}
\end{center}
\caption{\footnotesize
The same as Fig. \ref{fig3} but $e_1=0.500$ and $e_2=0.495$.
}
\label{fig4}
\end{figure}

In Fig. \ref{fig3}, the eigenvalues  $\ce^{PT}$  and  $\ce^{'PT}$
of (\ref{ham2o1}), left panel, and (\ref{ham2o2}), right panel, are shown. 
The corresponding eigenfunctions   shown in the lower part of
Fig. \ref{fig3}.  As can be seen from the results, the 
level repulsion appearing between the two EPs is
accompanied by a complete (1:1) mixing of the eigenfunctions. 
The mixing vanishes only far from the EP (Figs. \ref{fig3}.e and f). 

This result is in full analogy to 
the results shown in Figs. \ref{fig1}.e,f and  \ref{fig2}.e
for open quantum systems with imaginary $\omega$ where  width bifurcation
is accompanied by a complete mixing of the eigenfunctions between the
two EPs; and the mixing vanishes only far from the EPs. 
Further numerical studies have shown that also the phases of the
eigenfunctions always jump  by $\pi /4$ at the EPs (not shown in
Fig. \ref{fig3}).

We state therefore the following.
The results of Fig. \ref{fig3} obtained from calculations 
for systems with gain and loss and with real $w$ are formally similar to
those received for open quantum systems with imaginary coupling
coefficients $\omega$ (lower row in Figs. \ref{fig1} and \ref{fig2}).
In the two cases, the role of energy and width is formally exchanged. 

In order to receive a better understanding of the role of gain 
in Fig. \ref{fig3}, we performed
another calculation with slightly different energies $e_i$ of the two
states. The results shown in Fig. \ref{fig4} are very similar to those
in Fig. \ref{fig3}. The differences are of the same type as those
obtained in corresponding 
calculations for open quantum systems with $\omega =0.05 i$, see 
Fig. 1 (left panel) in \cite{elro3} with $\gamma_1 = \gamma_2$  and 
Fig. 2 (left panel) in \cite{elro3}
with $\gamma_1 \approx \gamma_2$, respectively.

\begin{figure}[ht]
\begin{center}
\includegraphics[width=13cm,height=13cm]{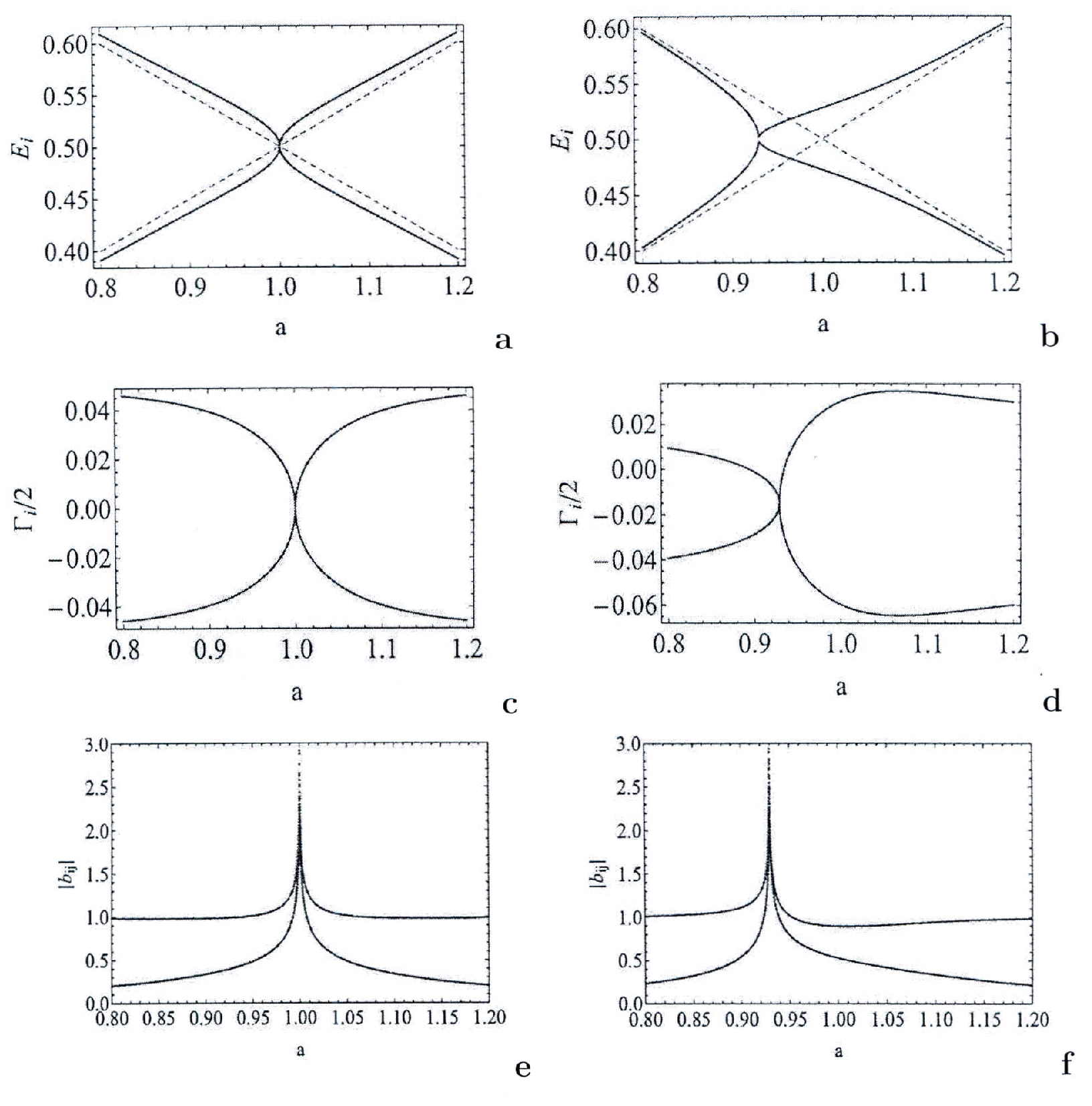}
\vspace{-.6cm}
\end{center}
\caption{\footnotesize
Energies $E_i$ (top),  widths $\Gamma_i/2$ (mid)
and mixing coefficients $|b_{ij}|$ (bottom) 
of  $N=2$  states of an open  quantum system with gain and loss which 
is coupled to one common channel, as a function of  $a$. 
The parameters are 
$e_{1}=1-a/2; ~e_2=a/2$ and
$\gamma_1/2=-0.05; ~\gamma_2/2=0.05; ~\omega = 0.05$ (left panel);
~$\gamma_1/2=-0.05; ~\gamma_2/2=0.0205; ~\omega = 0.05\,(1+i)/\sqrt{2}$ 
(right panel)
}
\label{fig11}
\end{figure}

Finally, we perform calculations with the Hamiltonian 
(\ref{form1}) but different signs for the two $\gamma_i$ (and
$\omega =\omega_{12} =\omega_{21}$). In this case, the eigenvalues  
$\ce_{i,j} \equiv E_{i,j} + \frac{i}{2} \Gamma_{i,j}$ are given by
(\ref{int6}) with 
\begin{eqnarray}
Z  =  \frac{1}{2} \sqrt{(e_1 - e_2)^2 -\frac{1}{4}(\gamma_1 - \gamma_2)^2
 +i\, (e_1-e_2)(\gamma_1 - \gamma_2) + 4 \omega^2} \; . 
\label{sign}
\end{eqnarray}
According to the condition $Z=0$ for the appearance of an EP, we have one EP 
at the crossing point  $a=a_{\rm cr}$ of the two $e_i$ trajectories
(where  $e_1(a)=e_2(a)$), if $\gamma_1 = -\gamma_2$ is parameter independent 
and $\omega = |\gamma_i /2|$ is real. There is however no EP 
when $\omega$ is imaginary. 
If $\omega $ is complex and the widths $\gamma_i$ of the two states have
different signs, there is also one EP. 
We show the corresponding numerical results with one EP in Fig. \ref{fig11}.   

We underline here that the results of  Fig. \ref{fig11} are obtained
by using the Hamiltonian (\ref{form1}) for a system with parameter
independent values of loss and gain. As usual, the EP appears at the
crossing point of the energy trajectories if $\omega$ is real. 
The system  shows  the characteristic 
features of an open quantum system. A balance between gain and loss
may appear, is however not necessary.
Systems of this type will  surely allow many different applications.

\section{Conclusions}
\label{con}

The results presented in the present paper show clearly the 
common features as well as the main 
difference between Hermitian and non-Hermitian quantum physics
when describing small systems coupled to a small number of
well-defined decay channels. 
Far from the singular EPs in non-Hermitian quantum physics, 
everything is analytical as in Hermitian quantum physics:  
Fermi's golden rule holds and counterintuitive results do not occur;
the eigenfunctions of the Hamiltonian are nearly orthogonal;
and the differences 
between Hermitian and non-Hermitian quantum physics practically 
vanish. At (and near to) EPs, however, the functional change of 
the dependence of the observables changes radically. It
is {\it non-analytical} and Fermi's golden rule does 
{\it not} hold. Instead, so-called counterintuitive results appear. 
This happens under the influence of the environment which 
is extremely large in the neighborhood of EPs
where the eigenfunctions of the Hamiltonian 
are really biorthogonal. The environment itself
represents an {\it infinitely large number of degrees of freedom} 
(continuum of scattering wavefunctions). It can be changed by means
of external forces, however it can never be deleted from an open
quantum system. Since all the individual states of the system are 
coupled to the common environment, their wavefunctions become 
mixed due to this coupling. Although this is a second-order process, 
it becomes the dominant one near to an EP.
 
This conclusion is based on the analytical and numerical results 
shown and discussed in the present paper. On the one hand, 
the differences between  
calculations with Hermitian and non-Hermitian Hamilton 
operator almost vanish far from EPs, see Fig. \ref{fig1a}.d. 
On the other hand,  
counterintuitive results determine the dynamics of the system
in the neighborhood of EPs. Most visible (and known for quite 
a long time) is the reduction of the lifetime of one of the 
two neighboring states in spite of increasing (imaginary) 
coupling strength between system and common
environment. This result originates  at the EP as 
all our calculations show.

The strong influence of an EP onto the dynamics of an
open quantum system can be expressed quantitatively
by the phase rigidity of the eigenfunctions of the non-Hermitian
Hamilton operator which is defined in (\ref{eif11}). The 
eigenfunctions are biorthogonal, and the 
phase rigidity vanishes in approaching an EP. Here,
the wavefunctions differ from one another 
by only a phase factor. Such a result is, of course, completely 
different from that what is known in Hermitian quantum physics. 
It explains why {\it the results obtained in non-Hermitian quantum 
physics differ substantially from those of Hermitian quantum 
physics only in the neighborhood of EPs}.  

The meaning of the environment for the physics of open quantum systems 
is confronted recently with existing experimental data 
in the review \cite{robi}. Further experimental and theoretical 
studies along the lines sketched in the present paper are necessary 
in order to receive more information on open quantum systems (which
are embedded into a common well-defined environment)
and to describe them by means of a non-Hermitian Hamilton operator.
The results are basic also for a better understanding of processes
occurring in optics, e.g. of the Dicke superradiance, as mentioned above.  
By choosing an appropriate
environment, it is possible to manipulate the system and to produce, by
doing this, systems with desired properties. These results are of 
importance for basic research as well as for applications.

\vspace{.5cm}

%\section*{References}

\newpage

\setcounter{section}{0}
\setcounter{equation}{0}
\setcounter{figure}{0}

\section*{\large \bf Part II: Three and more states}

\section*{Abstract}

Using the formalism for the description of open quantum systems 
by means of a non-Hermitian Hamilton operator, we study 
the occurrence of {\it dynamical phase transitions} 
as well as their relation to the singular exceptional points (EPs). 
First, we provide the results of an analytical study for the 
eigenvalues of three crossing states. These crossing points are of
measure zero. Then we show numerical results 
for the influence of a nearby ("third") state onto an EP. 
Since the wavefunctions of the two crossing states are mixed in a finite 
parameter range around an EP, three states of a physical system
will never cross in one point.
Instead, the wavefunctions of all three states are mixed in a finite 
parameter range in which the ranges of the influence of different EPs
overlap. We may relate these results  to  dynamical phase 
transitions observed recently in different experimental studies. 
The states on both sides 
of the phase transition are non-analytically connected.

\section{Introduction}
\label{intr}

Phase transitions are studied since very many years in different physical 
systems. In spite of their special features characteristic of every system,  
common to all of them is that the states at one side of the transition are
not analytically connected to those at the other side. Recently, phase 
transitions are observed experimentally and investigated theoretically also
in small open quantum systems. Mostly, the results are counterintuitive,
when considered from the point of view of Hermitian quantum physics. 
Some years ago, 
results of such a type could be explained 
qualitatively in the framework of non-Hermitian quantum physics by relating them to a phase 
transition occurring in an open quantum system \cite{Jumuro}. 
At the same time, these results are related \cite{Hemuro}
to the existence of singular points \cite{Kato}, 
the  so-called exceptional points (EP), emerging in the non-Hermitian formalism. 

By now, we have a much better understanding of these phenomena, 
especially in the field of mesoscopic physics where it is possible to trace 
the phase transition by means of parameter variation. A detailed discussion
can be found in the recent review \cite{Robi} where theory based on 
non-Hermitian quantum physics, is confronted with known experimental results
on open quantum systems. 
In recent papers, this type of phase transition is called mostly 
{\it dynamical phase transition (DPT)} in order to underline its relation
to the dynamics of open quantum systems. 
Much less understood are similar phenomena in optics although they are known experimentally for a long time, e.g. the Dicke superradiance \cite{Dicke}.

As a result of many differnt studies
during last years (for references see \cite{I}), the singular EPs 
of the non-Hermitian formalism play really an important role for the occurrence  of phenomena called counterintuitive in the Hermitian formalism. 
In \cite{I}, the
non-Hermitian formalism is sketched. It
allows us to study not
only the eigenvalues but also the eigenfunctions of a non-Hermitian
Hamilton operator $\ch$ in the neighborhood of 
the EPs.

At an EP, two eigenvalues of the Hamiltonian $\ch$ coalesce 
while the corresponding eigenfunctions differ from one another
only by a phase. Such a situation is possible only beyond standard  
Hermitian quantum physics since the eigenfunctions of {\it all} states 
of a Hermitian Hamilton operator are orthogonal (also when their 
eigenvalues accidentally coalesce), while those of a non-Hermitian operator are biorthogonal, and the phases of the eigenfunctions relative to one another may change.

In \cite{I},  analytical and numerical results
for the eigenvalues as well as for the eigenfunctions of a 
non-Hermitian Hamiltonian are provided;  
the influence of EPs onto physical observables is discussed; 
the appearance of nonlinear terms in the Schr\"odinger equation 
around EPs is considered; as well as their relation to the  
phases of the two eigenfunctions relative to one another.  
The reduction of the phase rigidity of the eigenfunctions near to an EP
means that the environment of scattering wavefunctions with its 
infinite large number of degrees of freedom plays here an extremely 
important role for the dynamics of the system since it mediates
a coupling of the different states of the system via the environment. 
As a consequence, the wave functions of the two crossing states are
mixed (via the environment) in a {\it finite} parameter range around 
an EP. They provide therefore valuable information on the environment 
not only at one point in the continuum (which is of measure zero),
but in a finite parameter range around this point. 

According to \cite{I}, the influence of EPs on 
the dynamical properties of open quantum systems consists in  
the following:
(i) The eigenvalues of $\ch$ show a non-analytical behavior (deviations
from Fermi's golden rule) at and in the vicinity of an EP. 
Width bifurcation occurs due to  
Im($\omega$), while level repulsion is caused by Re($\omega$); 
(ii) The phases of the eigenfunctions 
are not rigid in a finite neighborhood of an EP, and the phase 
rigidity $r_i \to 0$ at the EP.  This dynamical feature
allows the environment to influence the
system extremely strong at and near to an EP.
These results hold true for open many-body quantum systems the states 
of which decay into the environment of scattering states,
as well for those systems which may absorb particles from the
environment.

The main interest of our present study on non-Hermitian 
quantum systems  is to  find an answer
to the question how  different EPs influence one another 
and how they are related to a DPT
occurring in a physical system. Most studies have to be performed 
numerically, what is an expression of the well-known fact that the
states below and beyond a phase transition are non-analytically connected. 
We restrict ourselves, in our paper, to the coherent (collective)
phenomena induced in an open quantum system by embedding it into a
common environment.  These phenomena are very robust. 
We will not consider decoherent phenomena in the present paper which
arise by coupling the system to a large nonspecific environment, 
since this coupling does not cause any global new features of the
system such as e.g. a DPT, in which we are interested.

In our paper we consider first the crossing of the eigenvalues of
three states  analytically and discuss shortly their relevance for physical
processes (Sect. \ref{more1}). In the following Sect. \ref{num3} we investigate
numerically the influence of a nearby state onto an EP. Here, we 
consider not only the eigenvalues but also the eigenfunctions of the 
non-Hermitian Hamilton operator. The numerical
studies are performed for open quantum systems with emission ({\it loss})
of particles into the environment; and also for systems which 
additionally may  absorb particles
from the environment ({\it systems with loss and gain}). 

In Sect. \ref{dyn} of the present paper, we address the problem of the relation 
of EPs to DPTs appearing in systems with more than two states. We are able to
justify the restriction of the parameter dependence of the Hamiltonian
$\ch$ to its non-Hermitian part what allows us to receive quickly
information on the most important global spectroscopic redistribution 
processes occurring in open quantum systems, i.e. on the main features
of the DPTs.  We discuss also numerical results obtained 
for systems with gain and loss and their relation to those obtained for open
quantum systems. Some  conclusions on the meaning of EPs for the
dynamics of open quantum systems are drawn in Sect. \ref{concl}.
Here, we point also to the meaning of the results for phenomena observed 
in optics, e.g. for the Dicke superradiance.

\section{Crossing of $N=3$ states in an open quantum system}
\label{more1}

We consider a system consisting of three levels  coupled to
one common continuum of scattering wavefunctions. The Hamiltonian
reads 
\begin{eqnarray}
\label{f1}
\ch^{(3)}=\left[
\begin{array}{ccc}
\epsilon _{1}=e_{1}+i\frac{\gamma _{1}}{2} & \omega _{12} & \omega _{13} \\
\omega _{21} & \epsilon _{2}=e_{2}+i\frac{\gamma _{2}}{2} & 0 \\
\omega _{31} & 0 & \epsilon _{3}=e_{3}+i\frac{\gamma _{3}}{2}
\end{array}
\right] 
\end{eqnarray}
where $\omega_{23}=\omega_{32}=0$ is assumed by using the doorway
  picture \cite{Elro3}.
 For simplicity, we assume that all coupling coefficients are equal to
 one another, $\omega _{ij}=\omega $.
They may be real or complex.

According to \cite{Bronstein}, the eigenvalues can be determined 
from the polynomial
\begin{eqnarray}
\label{f3}
\lambda ^{3}+R\lambda ^{2}+S\lambda +T=0 
\end{eqnarray}
with
\begin{eqnarray}
\label{f4}
R&=&-(\epsilon _{1}+\epsilon _{2}+\epsilon _{3}) \nonumber \\
S&=&\epsilon _{1}\epsilon _{2}+\epsilon _{1}\epsilon _{3}+\epsilon
_{2}\epsilon _{3}-2\omega ^{2} \nonumber \\
T&=&\omega ^{2}\epsilon _{2}+\omega ^{2}\epsilon _{3}-\epsilon _{1}\epsilon
_{2}\epsilon _{3}
\end{eqnarray}
Eq. (\ref{f3}) can be transformed to 
\begin{eqnarray}
\label{f9}
y^{3}+py+q=0
\end{eqnarray}
with
\begin{eqnarray}
\label{f10}
y&=&\lambda +\frac{R}{3}   \\
\label{f7}
p&=&\frac{3S-R^{2}}{3}  \\[.2cm] 
\label{f8}
q&=&\frac{2R^{3}}{27}-\frac{RS}{3}+T 
\end{eqnarray}
Using  (\ref{f7}), (\ref{f8}) and 
\begin{eqnarray}
\label{f15}
u&=&\sqrt[3]{-\frac{q}{2}+\sqrt{
\left(\frac{p}{3}\right)^{3}+\left( \frac{q}{2}\right) ^{2}}}\\[.2cm]
\label{f16}
v&=&-\frac{p}{3u}
\end{eqnarray}
the three eigenvalues read
\begin{eqnarray}
\label{f11}
\lambda _{1}&=&\frac{u+v}{2}-\frac{R}{3}\\[.2cm]
\label{f12}
\lambda _{2}&=&-\frac{u+v}{2}-\frac{R}{3}+\frac{u-v}{2}i\sqrt{3}\\[.2cm]
\label{f13}
\lambda _{3}&=&-\frac{u+v}{2}-\frac{R}{3}-\frac{u-v}{2}i\sqrt{3}
\end{eqnarray}
When $y=0$ and  $q=0$ the three eigenvalues cross 
according to (\ref{f10}) and (\ref{f9}). Here
\begin{eqnarray}
\label{f18}
\lambda _{1,2,3}=-\frac{R}{3} =
\frac{\epsilon _{1}+\epsilon _{2}+\epsilon _{3}}{3}
\end{eqnarray}
where the definition (\ref{f4}) of $R$ is used. Further, 
\begin{eqnarray}
\label{f25}
u=\sqrt{\frac{p}{3}} \; ; \quad
~v= -\sqrt{\frac{p}{3}}=-u \; .
\end{eqnarray}

Equation (\ref{f18}) shows that the crossing point is determined by the
distance $R$ between the three
complex energies $\epsilon_i$ of the states. The way 
on which the crossing point is approached plays therefore an important
role. When approaching the crossing point by means of $u=-v \to 0$,
the eigenvalues  (\ref{f11}) to (\ref{f13}) at the crossing point read
\begin{eqnarray}
\label{f27}
\lambda _{1}&=&-\frac{R}{3}\\[.2cm]
\label{f28}
\lambda _{2}&\to&-\frac{R}{3}+iu\sqrt{3}\\[.2cm]
\label{f29}
\lambda _{3}&\to&-\frac{R}{3}-iu\sqrt{3}
\end{eqnarray}
The two eigenvalues $\lambda_2$ and $\lambda_3$ form an EP when $u\to 0$.
The eigenvalue  $\lambda_1$ is not influenced by this
condition. It plays the role of an observer state that can be
exchanged at the EP with one of the other two states, but does not
participate in the spectroscopic redistribution processes taking place 
at the crossing point. When however $u=0$  and the crossing point is
approached by varying another parameter (being independent of $u$),
all three states participate in the spectroscopic redistribution processes. 

The conclusion from this analytical study is the following. 
According to  the parameter varied in approaching the crossing point
of three states, two different types of crossing points exist. 
The differences between these two types of crossing points  
consist in the following
\begin{enumerate}
\item[--]
when the crossing point is approached by $u \to 0$, 
two states show the signatures of an EP 
while the third state is an observer state that, generally, may exchange with
one of the other states;
\item[--]
when the crossing point is approached by keeping constant 
$u= 0$  and varying another independent parameter, 
the three states form together a common crossing point, at which all 
three states  participate in the spectroscopic redistribution processes.
\end{enumerate}
In any case, the spectroscopic redistribution caused by the crossing
of three states  is a nonlinear process according to (\ref{f3}).
The results hold for systems with  {\it loss} (all $\gamma_i$ negative) 
as well as for systems with {\it loss and gain} (negative and positive
$\gamma_i$).

The detailed study of the crossing point of all three eigenvalues of the 
Hamiltonian (\ref{f1}) is of {\it formal-mathematical} interest
because it is a point in the continuum and therefore of measure zero.
The geometric phase related to the crossing point of three states
is, in any case, different from that of an EP 
for the crossing of two  levels.  This
holds true even in the case when the third state is an observer state since
this state may  exchange with each of 
the two other states. The geometric phase depends  therefore 
on the number $N$ of crossing states -- a result that is difficult to
understand for a realistic physical system.

According to the results obtained and discussed in 
\cite{I}, any "third" state of a realistic physical system
will cross or avoid crossing with one 
of the two states in the neighborhood of the EP whose wavefunction 
differs however from that of the original state (because the wavefunctions
of both states are mixed with one another near to the EP). 
At each of the crossings, the corresponding geometric phase is well 
defined: it is the geometric phase of an EP related to two states,
and differs therefore from the Berry phase of a diabolic point by a
factor 2 (see \cite{Top}). 

While the properties of  isolated EPs are  studied experimentally as
well as theoretically in many papers, the influence of a nearby state 
onto an EP is not at all investigated, up to now. In this case, 
the areas of influence of different EPs are expected to overlap, and 
nonlinear terms will appear not only at the EPs themselves but, above all, 
in a {\it finite} vicinity of them.  Analytical studies are restricted  
due to these nonlinearities. When combined with numerical studies 
they are expected to give reliable results, see the next section.

\section{Influence of a nearby state onto an exceptional point}
\label{num3}

\subsection{Open quantum system with three nearby states}
\label{num3a}

\begin{figure}[ht]
\begin{center}
\includegraphics[width=13cm,height=13cm]{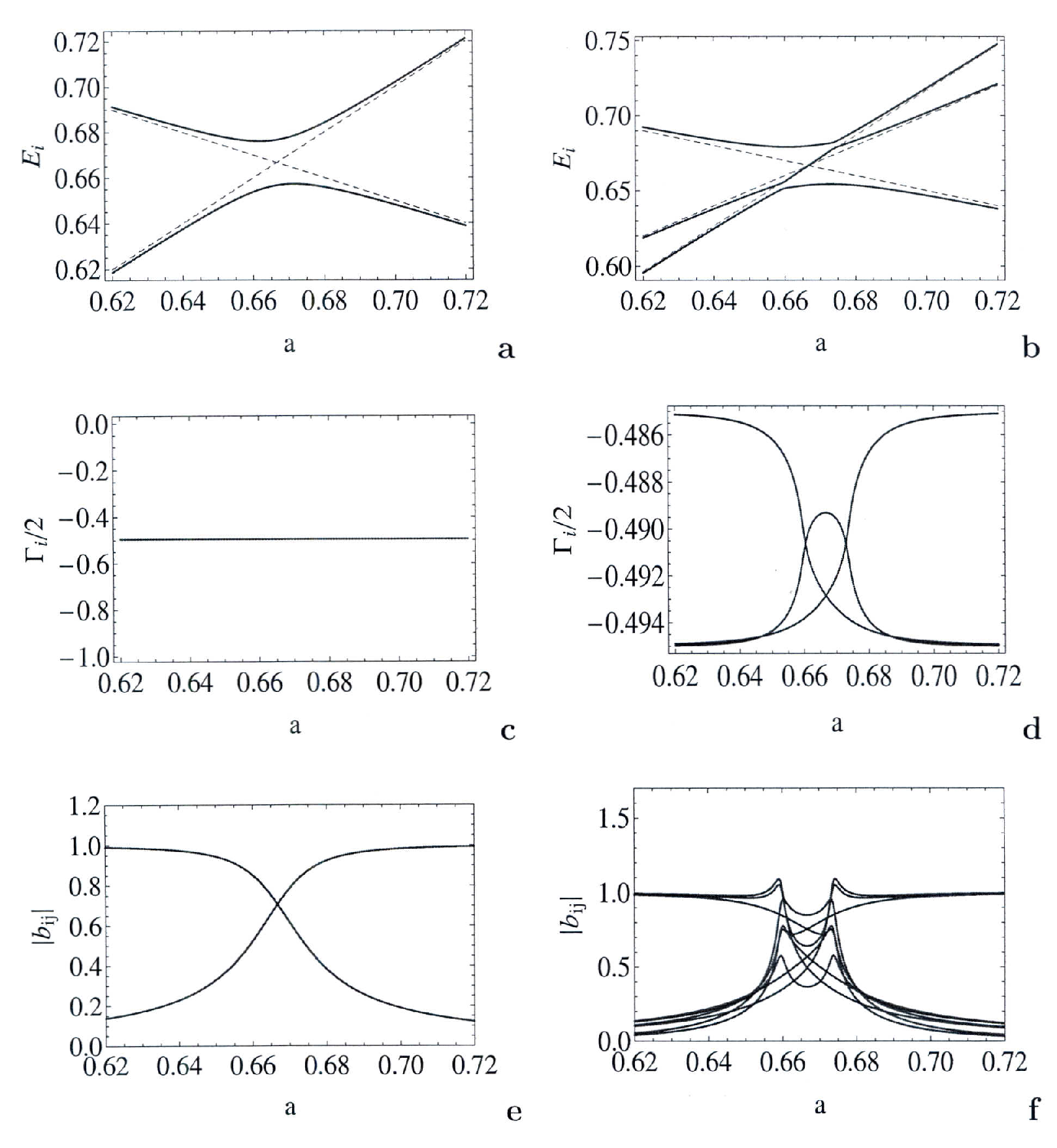}
\end{center}
\caption{\footnotesize
Energies $E_i$ (top), widths $\Gamma_i/2$ (mid) and  mixing
coefficients $|b_{ij}|$ (bottom) of the eigenfunctions $\Phi_i$  of
$N=2$ (left panel a,c,e) and $N=3$ (right panel b,d,f) states of an
open system
coupled to a common channel by $\omega = 0.01$ as a function of
$a$. Parameters: $e_1=1-1/2~a; ~~e_2=a$; ~~$e_3=-1/3+3/2~a$ (b,d,f);
 ~~$\gamma_1/2= \gamma_2/2 =- 0.495$; ~~$\gamma_3/2=- 0.485$ (b,d,f).
The dashed lines in (a,b) show $e_i(a)$.
}
\label{fig5}
\end{figure}

\begin{figure}[ht]
\begin{center}
\includegraphics[width=13cm,height=13cm]{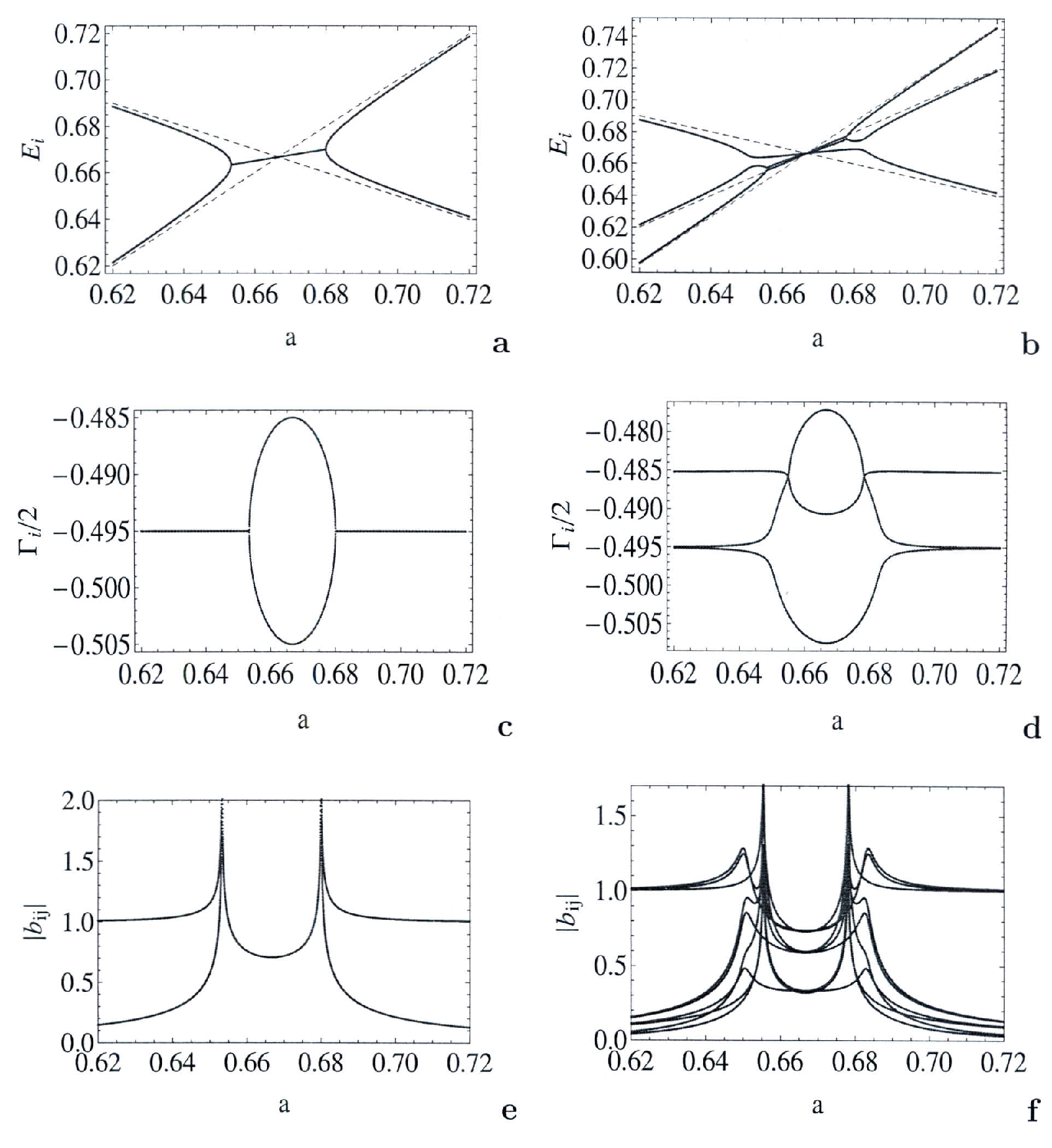}
\end{center}
\caption{\footnotesize
The same as Fig. \ref{fig5} but $\omega = 0.01 i$ and 
$\gamma_3/2=- 0.4853$ (b,d,f). 
}
\label{fig6}
\end{figure}

\begin{figure}[ht]
\begin{center}
\includegraphics[width=14cm,height=11cm]{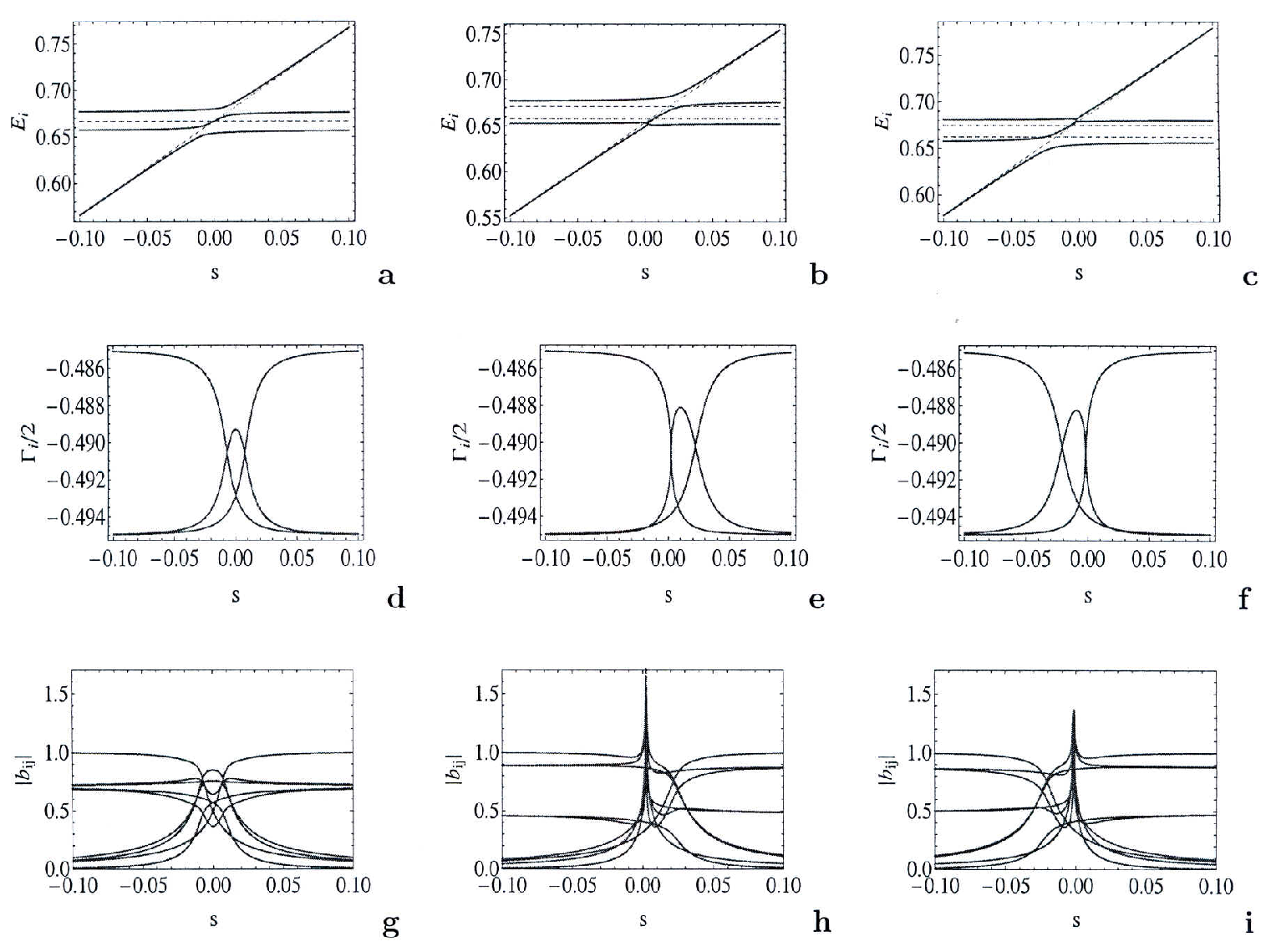}
\end{center}
\caption{\footnotesize
Energies $E_i$ (top),  widths $\Gamma_i/2$ (mid)
and mixing coefficients $|b_{ij}|$ (bottom) 
of $N=3$  states of an open system coupled 
to one common channel as a function of  $s$ with $e_3=s-1/3+3/2a$. 
~$\omega = 0.01$; ~$a=a_{cr}=2/3$ (left);  ~$a=a_{1}=0.6539<a_{cr}$ (mid);
 ~$a=a_{2}=0.675>a_{cr}$ (right). The $e_1,e_2,\gamma_1,\gamma_2,
 \gamma_3$ are the same as in Fig. \ref{fig5}.
The dashed lines in (a,b,c) show $e_i(s)$. 
}
\label{fig7}
\end{figure}

\begin{figure}[ht]
\begin{center}
\includegraphics[width=14cm,height=11cm]{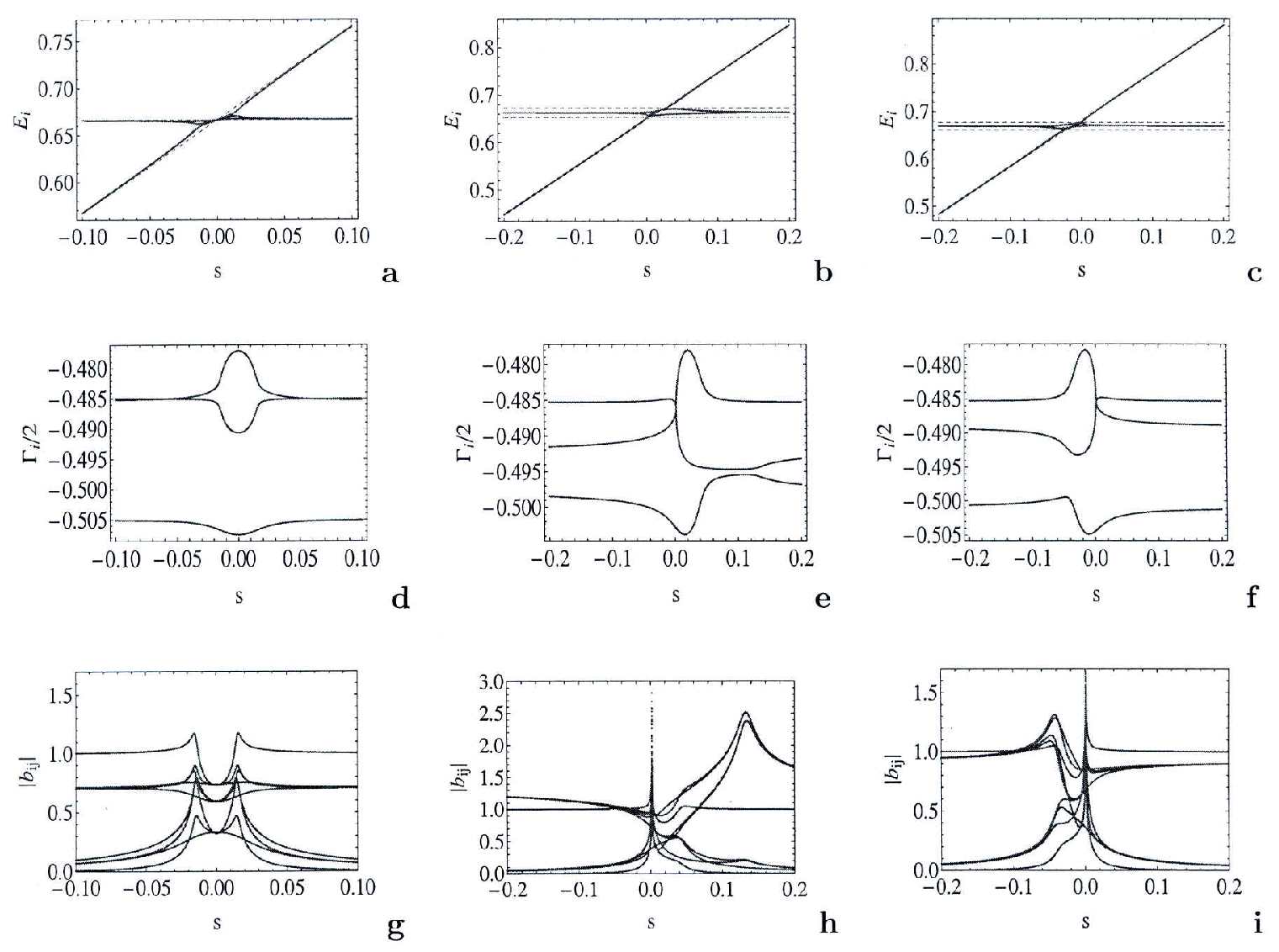}
\end{center}
\caption{\footnotesize
Energies $E_i$ (top),  widths $\Gamma_i/2$ (mid) 
and mixing coefficients $|b_{ij}|$ (bottom)
of $N=3$  states of an open system coupled 
to one common channel as a function of  $s$ with $e_3=s-1/3+3/2a$. 
~$\omega = 0.01i$; ~$a=a_{cr}=2/3$ (left);  ~$a=a_{1}=0.6539<a_{cr}$ (mid);
 ~$a=a_{2}=0.6774>a_{cr}$ (right). The $e_1,e_2,\gamma_1,\gamma_2,
 \gamma_3$ are the same as in Fig. \ref{fig6}.
The dashed lines in (a,b,c) show $e_i(s)$. 
}
\label{fig8}
\end{figure}

The results of numerical calculations shown in Figs. \ref{fig1}  
to \ref{fig8} are
performed with the non-Hermitian Hamiltonian (\ref{f1})
for an open quantum system and with Eq. (24) of \cite{I} for the
coupling coefficients $\omega$. We assume that only the energies $e_i$
of the states depend on a parameter $a$ while the widths $\gamma_i$ 
are constant in the considered parameter range.  
We show  the eigenvalues ${\cal E}_i = E_i
+ i/2~\Gamma_i$ and the mixing coefficients $|b_{ij}|$ of the
eigenfunctions $\Phi_i$ of (\ref{f1})  for real coupling coefficient
$\omega$ (Fig. \ref{fig5}) as well as for imaginary $\omega$
(Fig. \ref{fig6}) and compare them with those of $\ch^{(2)}$, Eq. (4) 
of \cite{I}, in the neighborhood of the EP.

In the left panel of Fig. \ref{fig5} (real $\omega$), we see the typical 
avoided crossing of two levels with an exchange of the wavefunctions
around the critical parameter value $a=a_{\rm cr} = 2/3$ (see Sect.
IID of \cite{I}). The right
panel of Fig. \ref{fig5} shows the crossing of the two levels with a third one.
Here, two intersections can be seen: the first one at
$a=a_1=0.65775<a_{\rm cr}$
and the second one at $a=a_2=0.675>a_{\rm cr}$. At the two intersections
$|b_{ij}|>1$  what is a clear hint to the existence of an EP. 
The eigenfunctions are mixed not only in the parameter
range between the two intersections but also in a comparable
large parameter range beyond them. As can be seen from the eigenvalue
pictures Fig. \ref{fig5}.b,d  the third state interacts with the two 
other ones and exchanges with them. 

Fig. \ref{fig6} shows the results obtained with imaginary $\omega$. 
In the left panel, we see two EPs at $a=a_1= 0.6539 < a_{\rm cr}$ and $a=a_2=
0.6774 > a_{\rm cr}$, respectively. The appearance of two EPs and the width
bifurcation between them is
characteristic for two levels having the same (or similar)
widths $\gamma_i$ (see Sect. IID of \cite{I}).  In the right panel of  
Fig. \ref{fig6}, a third level crosses the energy of the two
states in the parameter range in which the widths of the two states bifurcate. 
Width bifurcation occurs now  with participation of all three states
(Fig. \ref{fig6}.d).
Altogether, the width bifurcation is stronger than in the two-level
case. At $a=a_{\rm cr}$, the difference between the largest and
smallest values of
$\Gamma_i/2$ is 0.02 in the two-level case and 0.03 in the three-level
case. Here, the widths of the three levels appear in two groups: the 
largest width is much larger than  the two other widths which, on
their part, differ by a comparable small value from one another. 
This result is in agreement with   
the analytical result that the widths bifurcate at every
eigenvalue crossing.
 
As in the case of real coupling coefficients $\omega$
(Fig. \ref{fig5}), the eigenfunctions obtained with imaginary 
$\omega$ (Fig. \ref{fig6}) are mixed 
strongly in the parameter range between the two EPs as well as  
beyond this range. The mixing is symmetrical around the critical
parameter value $a=a_{\rm cr}$ in both cases 
and the large $|b_{ij}|$ point clearly to the existence of EPs.

In order to see more clearly the influence of a third level onto an
avoided level crossing or an EP, we keep fixed all the parameters used
in Figs. \ref{fig5} and \ref{fig6} with the exception of the energy 
$e_3$ of the third state. We choose $e_3=s-1/3+3/2~a$ and trace the  
eigenvalues and eigenfunctions of (\ref{f1}) as a function of $s$. 
At the value $a=a_{\rm cr}=2/3$, the three levels with energies
$e_i$ cross (see Figs. \ref{fig5}.b and \ref{fig6}.b).

We show the results  for different $a$ in Figs. \ref{fig7} (real $\omega$)
and \ref{fig8} (imaginary $\omega$).
The energies $E_i$ show level repulsion when $\omega$ is real,
while the widths $\Gamma_i$ of all three states bifurcate when
$\omega$ is imaginary. The eigenfunctions $\Phi_i$ are mixed strongly nearby
the intersection points and the EPs, respectively. Some
mixing remains  at large values of $|s|$.
The eigenvalue and eigenfunction pictures are completely symmetric
around $s=0$ when $a=a_{cr}$ while this is not the case when $a\ne a_{\rm cr}$, 
neither for real nor for imaginary $\omega$. These last cases are, of
course, more realistic than the symmetric one 
and will appear in physical systems. Comparing the results
for different $a$, one sees the sensitive
parameter dependence of the results in the critical region. 

All the results shown in Figs. \ref{fig5} to \ref{fig8}, are obtained
for the case that the widths of two states are equal, $\gamma_1 =
\gamma_2$. Very similar results are obtained when $\gamma_1 \approx
\gamma_2$. Further, the results of calculations with complex $\omega$ 
(not shown) show  characteristic features of the calculations
with  real as well as of those with imaginary $\omega$  
(similar as in Figs. 1 and 2 in \cite{I}).

\subsection{Three nearby states in a system with loss and gain}
\label{num3b}

\begin{figure}[ht]
\begin{center}
\includegraphics[width=13cm,height=13cm]{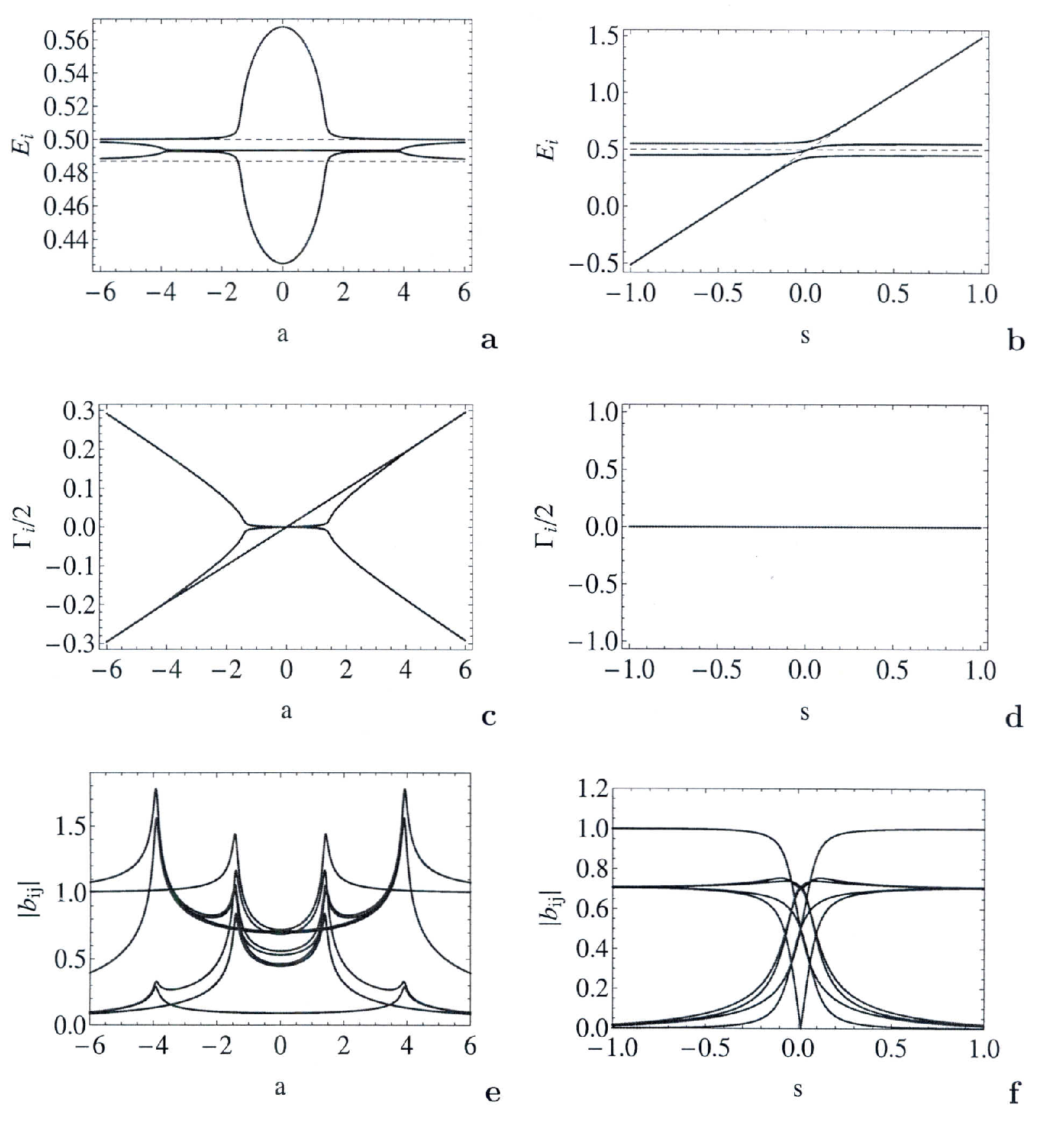}
\end{center}
\caption{\footnotesize
Energies $E_i$ (top),  widths $\Gamma_i/2$ (mid)
and mixing coefficients $|b_{ij}|$ (bottom) 
of $N=3$ states of a system with loss and gain, which is coupled 
to one common channel, as a function of  $a$ (left panel) and $s$
(right panel). 
The parameters are $e_{1}=e_2=0.5$ and $ e_3=0.487$ (left panel),
$e_3=0.487+s; ~a=0$ (right panel);
~$\gamma_1/2=-0.05a; ~\gamma_2/2=0.05 a; ~\gamma_3/2=0.05 a$; ~~ $w = 0.05$.
The dashed lines in (a,b) show $e_i(a)$. 
}
\label{fig9}
\end{figure}

\begin{figure}[ht]
\begin{center}
\includegraphics[width=13cm,height=13cm]{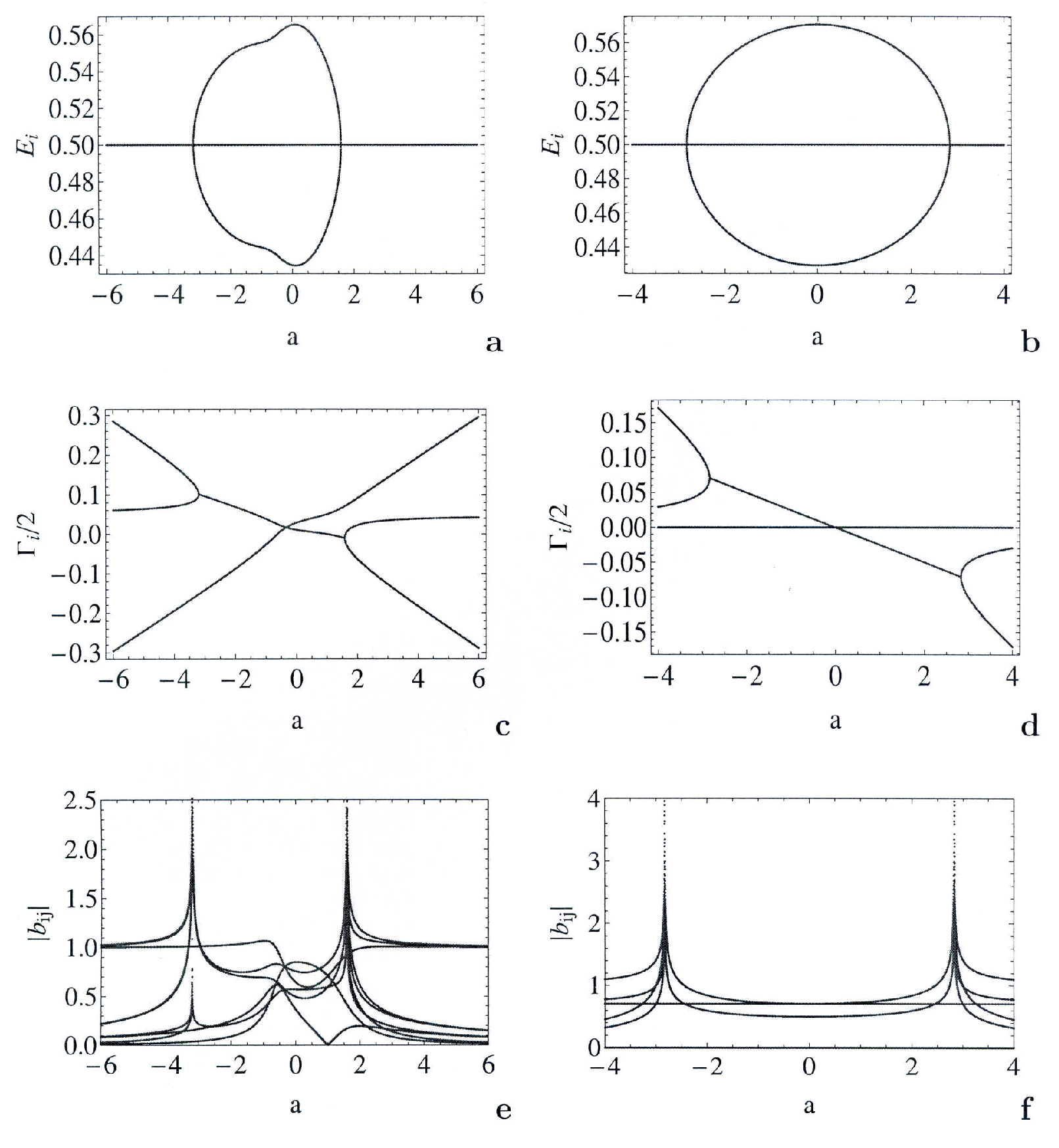}
\end{center}
\caption{\footnotesize
Energies $E_i$ (top),  widths $\Gamma_i/2$ (mid)
and mixing coefficients $|b_{ij}|$ (bottom) 
of  $N=3$  states of a system with loss and gain, which is coupled 
to one common channel, as a function of  $a$. 
The parameters are $e_{1}=e_2=e_3=0.5$ and
$\gamma_1/2=-0.05a; ~\gamma_2/2=0.05 a; ~\gamma_3/2=0.05$ (left panel);
~~$\gamma_1/2=-0.05 a; ~~\gamma_2/2=\gamma_3/2=0$ (right panel);
~~ $w = 0.05$.
}
\label{fig10}
\end{figure}

In Figs. \ref{fig9} and \ref{fig10} we show the influence of a 
nearby state onto an EP when the system has gain as well as loss
and the coupling coefficient $w$ is real. 
Instead of  width bifurcation occurring in the case of 
an open quantum system and imaginary coupling coefficient $\omega$, we see
now the separation of the states from one another in energy. However,
the third state has a large influence on the eigenvalue trajectories
as well as on the parameter range in which the eigenfunctions of $\ch$
are mixed.

In Fig. \ref{fig9} the widths $\gamma_i$ of the states $2$ and $3$
increase with $a$, while  that of state $1$ decreases with $a$. 
Due to this, there are altogether four EPs, 
at  $a\approx -4; ~-2; ~+2; ~+4$. At $a\approx -4$,
the widths $\Gamma_i$
of the two states with $\gamma_i \propto a$ separate from
one another while their energies $E_i$  coalesce. The opposite happens
at $a\approx -2$: the width  $\Gamma_i$ of the state with  $\gamma_i
\propto -a$ coalesces with that of one of the two states with    
$\gamma_i \propto a$ while the two states separate in energy at these
critical $a$ value.  The figure is symmetrical around $a=0$. The
eigenfunctions of $\ch$ are mixed in a parameter range, which is much
larger than the range of mixing without the third state.

In the right panel of Fig. \ref{fig9}, the influence of the third state
onto the eigenvalues and eigenfunctions  of $\ch$ is shown by varying
its distance $s$ to the two crossing states. The calculation is 
performed with $a=0$ and the results are symmetrical around $s=0$. 

In Fig. \ref{fig10}, we show results for the case that the width of
the third state is independent of the parameter $a$. It is 
$\gamma_1/2 = - \gamma_2/2 = - 0.05 a, ~\gamma_3/2=0.05 $ (left
panel) and $\gamma_1/2 = -0.05 a; ~\gamma_2/2=\gamma_3/2=0$ (right
panel). The figures are similar to one another. In
both cases, we have two EPs. 
The wavefunctions are strongly mixed between the two
EPs as well as in a finite parameter range beyond them.

\section{ Crossing of three (and more) states and dynamical phase transitions}
\label{dyn}

In the foregoing sections of the present paper, we showed
the influence of a nearby state onto an EP. First we considered
the problem analytically for three different eigenvalues 
without paying attention to the corresponding eigenfunctions
(Sect. \ref{more1}). 
As a result, the eigenvalues of all three states may coalesce in one
point either by forming an EP (with all its signatures) of two of the
states while the remaining state is uninvolved;
or by forming a new common singular point. In both cases, the
geometric phase differs from that of an EP since  the
states can be exchanged, even the
``uninvolved'' state may exchange with one of the other two states. 
A similar result for the geometric phase has
been obtained in \cite{Eva}. We found further that it depends on the
manner the crossing point is approached which of the two possibilities
will be realized.

However, these results cannot give any answer to the question how
different EPs influence one another. More important
than the crossing points themselves is  their influence onto the system
properties in some {\it finite} parameter range around an EP. All our 
calculations (Figs. \ref{fig5} to \ref{fig10}) have shown the 
generic feature of three nearby states\,: {\it the areas of influence 
of different EPs overlap and amplify, collectively, their impact onto 
physical values.} The eigenfunctions $\Phi_i$
are mixed in the basic wavefunctions $\Phi_j^0$  (see Eq. (23) in \cite{I}) 
in a finite critical parameter range. The range
is larger than in the 2-level case. Also the width bifurcation 
(when the $\omega$ are imaginary and  $\gamma_i < 0$) is
larger in the 3-level case than in the 2-level case (Fig. \ref{fig6}).
It occurs stepwise and may explain, in this manner,
the hierarchical trapping of resonance
states found many years ago \cite{Isrodi}.

The conclusion from our numerical studies is that,
in  systems with three states, the various states lose their
individual character in some critical parameter range, over which 
the distinct regions of influence of the various EPs overlap.
The same holds true when the number of states is larger than three, as
further numerical studies for four and more states have shown.
The significance of the individual EPs is therefore lost
in a finite parameter range, so that {\it more than two states 
of a realistic physical system are unable to coalesce at a
single point} (remember that an EP is a point in the continuum and 
its influence on physical observables can be seen {\it only} in its 
vicinity). While in the case of an isolated EP the two states are
exchanged at the EP (in accordance with the predictions of Eq. (18) in
\cite{I}), in the case with overlapping areas of influence of EPs,
the states are no longer directly related to the original
ones. Instead a spectroscopic redistribution, caused by several EPs,
takes place that is aimed at achieving a {\it dynamical stabilization of
the system} by accumulation  as much coupling strength between system 
and environment as possible onto just one specific state (in the
one-channel case) of the localized system. At the same time, the
remaining states decouple strongly from the environment.
The phase rigidity of the eigenfunctions is found to be reduced over 
a relatively large parameter range, where also significant
nonlinearities  appear. As mentioned already above, these
effects persist when the number of interacting states
is increased beyond three. It should be underlined here that this 
process of stabilization of the system as a whole occurs with 
the participation of {\it all} states and that, in the two-level case,
it corresponds to nothing but width bifurcation.
 
In any case, the process of stabilization is directly related to
the influence of the environment, with its infinitely large number 
of degrees of freedom, on the system at EPs. It may be related
therefore to a DPT (see \cite{Jumuro}). The relation between DPTs and
EPs can be seen clearly only in
the two-level case involving just a single EP. In the many-level case,
the role of the EPs is somewhat hidden because the areas influenced by
different EPs overlap and the positions of the EPs cannot be determined
analytically. The numerical calculations show, however, clearly that they
are responsible for the spectroscopic reordering processes which 
finally cause the DPT.  This can be seen in Figs. \ref{fig5} and
\ref{fig6}  in which the eigenvalues and eigenfunctions of $\ch$ are 
compared for the three-level case with those for the much simpler 
two-level case. According to these results, the non-Hermitian 
Hamilton operator  of the open quantum system can be 
approximated, in the region of the DPT,  quite well by
\begin{eqnarray} 
 \ch =  H^B -i\alpha VV^+
\label{ham_approx}
\end{eqnarray}
where the non-Hermitian perturbation $VV^+$ stands for the residuum, 
Eq. (3) of \cite{I}, and the principal value integral, Eq. (2) of \cite{I},  
is neglected (since its value is small when many states are nearby,
what is the case in the range over which the DPT occurs). 
Many calculations are performed, indeed,
by using the approximation (\ref{ham_approx}) and tuning the parameter
$\alpha$. The common feature of   
open quantum systems is that their quantum dynamics exhibits
{\it non-analytically connected states on either side of the transition}.

Summarizing, our results obtained for $N>2$ decaying
states (with ``loss'')  allows us to state 
that {\it the basic features of a DPT can be seen
already at a relatively small number of nearby states. 
A DPT is caused by several EPs (the crossing points of 
two resonance states) the areas of influence of which overlap. Fermi's
golden rule holds only far from the DPT.   }

Similar results hold for systems in which the environment allows not
only loss of particles but also gain (absorption of particles from the
environment). The main difference to a natural system  with only loss
of particles (corresponding to an emission into the environment) 
is, obviously,  that the coupling 
strength to the environment gives an additional real (Hermitian) part to
the non-Hermitian Hamilton operator. This part causes level repulsion
over a finite parameter range, and a balance between loss and gain
becomes possible.

\section{Conclusions}
\label{concl}

Concluding we state the following. The present-day high resolution
experimental studies require a description of quantum systems by
taking into account their embedding into the common continuum of 
scattering wavefunctions. This natural
environment exists always. It can be changed by means of external
fields, however it can never be deleted. 
This basic assumption of the description of open quantum systems, 
used in the present paper, is proven experimentally, see for example
\cite{Birdprx} where it is shown that
the interaction between two remote quantum  states 
in semiconductor nanostructures (quantum point contacts)
is essentially mediated by the continuum. The coupling of the open quantum
system to the environment causes the Hamiltonian to be non-Hermitian whose
eigenvalues $\ce_i$ are complex, generally. They provide not only the
energies $E_i=$ Re($\ce_i$) of the states of the system but also their lifetimes
which are inverse proportional to the widths $\Gamma_i/2=$ Im($\ce_i$). 
The feedback of the common environment onto the system is involved in the 
non-Hermitian Hamiltonian $\ch$ as well as 
in its eigenvalues  $\ce_i$ and eigenfunctions $\Phi_i$.
For distant levels, the feedback can be neglected, to a good approximation, 
and the non-Hermitian Hamiltonian $\ch$  passes smoothly into the
standard Hermitian Hamiltonian $H^B$ (for details see \cite{I}).

The coupling of the system to the common environment of scattering
wavefunctions entails some non-trivial mathematical problems.
First of all, there are singular points (EPs) in the continuum
at which the influence of the environment onto the system is extremely
large and which therefore 
influence strongly the dynamics of open quantum systems. They
cause nonlinear terms in the Schr\"odinger equation and non-rigid
phases of the wavefunctions, 
as shown and discussed in ref. \cite{I} for isolated EPs.
Different EPs influence each other when the ranges of their
interaction overlap. Here, a clustering of EPs occurs and causes a
strong mixing of the  wavefunctions of all states  in a finite
parameter range.   
According to the numerical results obtained for three states, the 
eigenvalues of all the three eigenstates of $\ch$ will {\it never cross exactly 
in one point} in a physical system. Another important feature is that
the environment can put its information 
on an infinite number of degrees of freedom 
(continuum of scattering wavefunctions) at the EP into the system. 
This mechanism is strengthened when the system has more than two
states as shown in the present paper. This mechanism
is aimed at the {\it stabilization of the system} by 
accumulation almost the whole coupling 
strength between system and environment onto one state 
(in the one-channel case). EPs may cause therefore DPTs with 
non-analytical connected states on both sides of the DPT. 
This statement corresponds to the results obtained qualitatively  
some years ago \cite{Jumuro}  for the appearance of a phase 
transition in an open quantum system described by a non-Hermitian 
Hamilton operator.

The results shown and discussed in the present paper are
obtained for open quantum systems described by a Schr\"odinger equation 
while many experimental results on DPTs are known in optics. An
example is the well-known Dicke superradiance \cite{Dicke} and the
recent studies on systems with loss and gain that started with 
\cite{Ptexp1,Ptexp2,Ptexp3} and are continued in many different 
investigations, e.g. \cite{Ptexp4,Ptexp5,Ptexp6}. In the studies 
with gain and loss, the equivalence of  
the optical wave equation and the quantum mechanical Schr\"odinger 
equation \cite{Equivalence1,Equivalence2,Equivalence3} 
(which holds true at least under special 
conditions) is used to explain the experimental results by means of 
a phase transition occurring under the influence of EPs. 
The results of the present paper are expected therefore to  be 
applicable also to the description of optical phenomena. 
Of special interest is the open question whether the Dicke 
superradiance \cite{Dicke} known since many years, is a DPT of the
type discussed in the present paper. In \cite{Elro3}, it
has been shown  that the transition from 
Autler-Townes-splitting to electromagnetically induced transparency 
in optics might be understood 
as a DPT between two very different processes taking
place as a function of the Rabi frequency. Other studies on a DPT 
and EPs in optics are performed  very recently \cite{Ramy,Domokos}. 

In any case, the results of the present paper will  allow us, on the
one hand, to receive a better understanding of the properties of 
open quantum systems and of DPTs. On the other hand, they will allow 
us to design new devices with  desired properties.

\vspace{1cm}

%\section*{References}


\begin{thebibliography}{99}
\bibitem{comment1} The coalescence of two eigenvalues of a
  non-Hermitian operator should not be confused with the degeneration
  of two eigenstates of a Hermitian operator. The eigenfunctions of two 
degenerate states are  different and orthogonal while those of two
coalescing states are biorthogonal and differ only by a phase, see Eqs. 
(\ref{int3}) to (\ref{eif5}).
\bibitem{comment2}
In difference to \cite{elro2}, the definitions
  $\epsilon_i = e_i + \frac{i}{2} \gamma_i $ and $\ce_i = E_i +\frac{i}{2}
  \Gamma_i$  (with $\gamma_i \le 0$ and $\Gamma_i \le 0$ for decaying states)
  are used in  the present paper
\bibitem{comment} 
 In studies of other researchers, the factor $i$ in 
(\ref{eif5}) does not appear. This difference is discussed 
and compared with experimental data in the Appendix of \cite{fdp1} and
in Sect. 2.5 of \cite{top}, see also Figs. 4 and 5 in \cite{berggren}.
\end{thebibliography}
\end{document}